\documentclass[11pt]{article}
\renewcommand{\arraystretch}{1.3}
\catcode`\@=11
\def\marginnote#1{}

\newcount\hour
\newcount\minute
\newtoks\amorpm
\hour=\time\divide\hour by60
\minute=\time{\multiply\hour by60 \global\advance\minute by-\hour}
\edef\standardtime{{\ifnum\hour<12 \global\amorpm={am}%
        \else\global\amorpm={pm}\advance\hour by-12 \fi
        \ifnum\hour=0 \hour=12 \fi
        \number\hour:\ifnum\minute<10 0\fi\number\minute\the\amorpm}}
\edef\militarytime{\number\hour:\ifnum\minute<10 0\fi\number\minute}

\def\draftlabel#1{{\@bsphack\if@filesw {\let\thepage\relax
      \xdef\@gtempa{\write\@auxout{\string
          \newlabel{#1}{{\@currentlabel}{\thepage}}}}}\@gtempa \if@nobreak
    \ifvmode\nobreak\fi\fi\fi\@esphack} \gdef\@eqnlabel{#1}}
    \def\@eqnlabel{}
\def\@vacuum{}
\def\draftmarginnote#1{\marginpar{\raggedright\scriptsize\tt#1}}

\def\draft{
  \oddsidemargin -.5truein
  \def\@oddfoot{\footnotesize \sl preliminary draft \hfil
    \rm\thepage\hfil\sl\today\quad\militarytime}
  \let\@evenfoot\@oddfoot \overfullrule 3pt
    \let\label=\draftlabel
    \let\marginnote=\draftmarginnote
  \def\@eqnnum{(\theequation)\rlap{\kern\marginparsep\tt\@eqnlabel}%
    \global\let\@eqnlabel\@vacuum}

  }
\makeatletter
\newdimen\normalarrayskip              % skip between lines
\newdimen\minarrayskip                 % minimal skip between lines
\normalarrayskip\baselineskip
\minarrayskip\jot
\newif\ifold             \oldtrue            \def\new{\oldfalse}
\def\arraymode{\ifold\relax\else\displaystyle\fi} % mode of array entries
\def\eqnumphantom{\phantom{(\theequation)}}     % right phantom in eqnarray
\def\@arrayskip{\ifold\baselineskip\z@\lineskip\z@
     \else
     \baselineskip\minarrayskip\lineskip2\minarrayskip\fi}
\def\@arrayclassz{\ifcase \@lastchclass \@acolampacol \or
\@ampacol \or \or \or \@addamp \or
   \@acolampacol \or \@firstampfalse \@acol \fi
\edef\@preamble{\@preamble
  \ifcase \@chnum
     \hfil$\relax\arraymode\@sharp$\hfil
     \or $\relax\arraymode\@sharp$\hfil
     \or \hfil$\relax\arraymode\@sharp$\fi}}
\def\@array[#1]#2{\setbox\@arstrutbox=\hbox{\vrule
     height\arraystretch \ht\strutbox
     depth\arraystretch \dp\strutbox
     width\z@}\@mkpream{#2}\edef\@preamble{\halign
\noexpand\@halignto
\bgroup \tabskip\z@ \@arstrut \@preamble \tabskip\z@ \cr}%
\let\@startpbox\@@startpbox \let\@endpbox\@@endpbox
  \if #1t\vtop \else \if#1b\vbox \else \vcenter \fi\fi
  \bgroup \let\par\relax
  \let\@sharp##\let\protect\relax
  \@arrayskip\@preamble}
\def\eqnarray{\stepcounter{equation}%
              \let\@currentlabel=\theequation
              \global\@eqnswtrue
              \global\@eqcnt\z@
              \tabskip\@centering
              \let\\=\@eqncr
 \halign to \displaywidth\bgroup
    \eqnumphantom\@eqnsel\hskip\@centering
    $\displaystyle \tabskip\z@ {##}$%
    \global\@eqcnt\@ne \hskip 2\arraycolsep
         $\displaystyle\arraymode{##}$\hfil
    \global\@eqcnt\tw@ \hskip 2\arraycolsep
         $\displaystyle\tabskip\z@{##}$\hfil
         \tabskip\@centering
    &{##}\tabskip\z@\cr}
\begingroup\ifx\undefined\newsymbol \else\def\input#1 {\endgroup}\fi
\newfont{\hr}{msbm10}
\newfont{\ams}{msam10}
\voffset= - 1.4in
\hoffset= - 1.0in         % switch off for draft style
\def\beq{\begin{equation}}
\def\eeq{\end{equation}}
\def\ba{\beq\new\begin{array}{c}}
\def\ea{\end{array}\eeq}
\def\be{\ba}
\def\ee{\ea}

\def\N2{${\cal N}=2$}

\def\1N{${\cal N}=1$}
\def\4N{${\cal N}=4$}
\def\nn{\nonumber}
\def\c{\check}
\def\p{\partial}

\title{{\bf
Unified description of correlators
\\
in non-Gaussian phases
of Hermitean matrix model
}
\vspace{.5cm}}
\author{{\bf A. Alexandrov}\thanks{E-mail: \ al@itep.ru}
\date{ } \\ {\small {\it MIPT} and
{\it ITEP, Moscow, Russia}}\\ \\
{\bf A. Mironov}\thanks{E-mail:
\ mironov@itep.ru; mironov@lpi.ac.ru}
\date{ } \\
{\small {\it Theory Department, Lebedev Physics Institute}
and {\it ITEP, Moscow, Russia}}\\ \\
{\bf A.Morozov}\thanks{E-mail: \ morozov@itep.ru}
\date{ } \\ {\small
{\it ITEP, Moscow, Russia}}}

\begin{document}

\vspace{1cm}

\maketitle

\vspace{-11.5cm}

\begin{center}
\hfill FIAN/TD-14/04\\
\hfill ITEP/TH-57/04\\
%\hfill hep-th/0412099
\end{center}

\vspace{10cm}

\begin{abstract}
Following the program, proposed in hep-th/0310113,
of systematizing known properties of matrix model partition functions
(defined as solutions to the Virasoro-like sets of linear differential
equations),
we proceed to consideration of non-Gaussian phases of the Hermitean
one-matrix model.
A unified approach is proposed for description of ``connected correlators"
in the form of the phase-independent ``check-operators" acting on
the small space of $T$-variables (which parameterize the polynomial $W(z)$).
With appropriate definitions and ordering prescriptions,
the multidensity check-operators look very similar to the Gaussian case
(however, a reliable proof of suggested explicit expressions in all loops is
not yet available, only certain consistency checks are performed).
\end{abstract}
\newpage
\tableofcontents
\newpage
\section{Introduction and definitions}

In \cite{amm1} we proposed to consider the matrix models partition
functions (a special class of $\tau$-functions subject to additional
set of {\it linear} differential equations
\cite{intmamo})
as the first family of special functions,
peculiar for the needs of the string theory, and started classification
and tabulation of their properties.
The natural place to begin is the Hermitean one-matrix model,
with the partition function ${\cal Z}(t)$ defined as a solution to the ordinary
Virasoro constraints \cite{virco}\footnote{
Sometimes they are called ``discrete" Virasoro constraints,
as opposed to ``continuum" ones \cite{contvir},
which give rise \cite{MMMKo} to the Kontsevich matrix model \cite{Ko,GKM}.
},
\be
\hat{\cal L}_m {\cal Z}(t) = 0, \ \ \ m\geq -1, \nn \\
\hat{\cal L}_m = \sum_{k=0}^\infty kt_k\frac{\partial}{\partial t_{k+m}} +
g^2\sum_{\stackrel{a,b\geq 0}{a+b = m}}
\frac{\partial^2}{\partial t_a\partial t_b}
\label{virc}
\ee

As explained in \cite{amm1}, the partition function ${\cal Z}(t)$ is
a sophisticated function of its infinitely many variables $t_0,t_1,\ldots$
and, as usual in theory of special functions, one is interested in
the two essentially different types of solutions to (\ref{virc}):

(i) formal series in powers of $t$-variables and

(ii) globally defined function of (at least, some) variables
$t_k$, which take values in some Riemann surface(s).

The usual situation is that the solutions of type (ii) provide a
kind of a linear basis in the space of all solutions of type (i),
and this provides some ground for the theory of ``analytical continuations"
or ``phase transitions" between different branches of partition function
(which are used to describe various phases of the related physical models).

In the case of the finite-size Hermitean one-matrix model, an interesting basis
of type (ii) is formed by the Dijkgraaf-Vafa partition functions
\cite{DV,DVfollowup}, which possess integral representations
in the form of the (specially defined) matrix integrals \cite{mamorep},
satisfy Givental-style decomposition formulas into multilinear
combinations of the Gaussian partition functions \cite{amm1},
and which already show some traces of emerging a
global definition in terms of Riemann surfaces \cite{globdef}.
A promising new step in this direction is recently made by Eynard
in \cite{Eynard}.
In the present paper we do not discuss these subjects and instead
concentrate on (i).
Then, as explained in \cite{amm1}, the interesting (though not exhaustive)
class of branches is specified by the three-step procedure:

(a) One requires existence of the ``genus expansion", i.e. request for
${\cal Z}(t)$ to depend on the scaling parameter $g^2$ in (\ref{virc}) as
\be
{\cal Z}(t) = \exp \left(g^{-2}{\cal F}(t;g)\right)
\ee
where the ``prepotential" ${\cal F}(t;g)$ is a formal series in {\it
non-negative} powers of $g^2$,
\be
{\cal F}(t;g) = \sum_{p=0}^\infty g^{2p} {\cal F}^{(p)}(t)
\label{gexpF}
\ee

(b) After a shift $t_k \longrightarrow T_k + t_k$, the partition
function and all the prepotentials
\be
{\cal Z}_W(t) = {\cal Z}(t-T), \nn \\
{\cal F}_W^{(p)}(t) = {\cal F}^{(p)}(t-T)
\ee
are formal series in {\it non-negative} powers of $t$-variables.
Different functions $W(z) \equiv \sum_{k=0} T_kz^k$ give rise to different branches,
the simplest ones being associated with polynomial $W(z)$.
In such cases, the degree $n$ and the roots $\alpha_i$ of
the derivative polynomial
\be
W'(z) = (n+1)T_{n+1}\prod_{i=1}^n (z-\alpha_i)
\label{rootsW'}
\ee
become important parameters, distinguishing between the phases.

In fact, as explained in \cite{amm1}, even after these two
steps of specifications, the branches are still not fully
separated, one more step is needed.

(c) The branches are fully specified by choosing
an {\it almost} arbitrary function
\be
F(T;g) \equiv {\cal F}_W(t=0;g) = \sum_{p=0}^\infty g^{2p}F^{(p)}(T)
\label{barpre}
\ee
of $T$- (or $\alpha$-) and $g$-variables: the "bare" prepotential.
The bare prepotential is constrained only by the {\it first two} reduced
Virasoro constraints,
\be
\check L_{-1} F(T,g) = \sum_{k=1}^{n+1} kT_k\frac{\partial F(T;g)}
{\partial T_{k-1}} = 0, \nn \\
e^{-g^{-2}F(T;g)}\check L_0 e^{g^{-2}F(T;g)} =
\sum_{k=1}^{n+1} kT_k\frac{\partial F(T;g)}
{\partial T_k} +
\left( \frac{\partial F(T;g)}{\partial T_0}\right)^2 +
g^2 \frac{\partial^2 F(T;g)}{\partial T_0^2} = 0
\label{redvirF}
\ee
which makes it an arbitrary function of $n$ variables, for instance, of
$T_0,\ldots,T_{n-1}$, and $g$.
All the correlators in a given phase depend on the choice of
$W(z)$ and of $F(T;g)$.

It is the purpose of the present paper to study these dependencies.
Ref.\cite{amm1}
contains the detailed description of the Gaussian branch, $n=1$,
with the (would be arbitrary) function $F(T_0;g)$ fixed to be
\be
\exp \left(g^{-2}F_G(T_0;g)\right) =
\frac{(g/T_2)^{-N^2/2}}{{\rm Vol}(SU(N))}
e^{-NT_0/g}
\label{Gprep}
\ee
Now we want to get rid of these restrictions and explain how
expressions for the Gaussian correlators can be generalized to the
phase with arbitrary $n$ and $F(T;g)$.
Significance of the Gaussian case is that the adequate quantities,
which provide a universal description of correlators in any phase,
the ``check-operator multidensities" $\check\rho^{(p|m)}$
(to be introduced in the next
section \ref{corf}) look practically the same as the Gaussian connected
correlators $\rho_G^{(p|m)}$, see Table 2.
Like the discussion of the Gaussian case in \cite{amm1},
this by itself does not provide an immediate
description of these correlators in terms of Riemann surfaces and
does not help to resolve the problems at level (ii),
but this step is the first one to make the systematic
description of non-Gaussian branches of the Hermitean one-matrix model.

We begin in s.\ref{corf} below with formulating the problem of evaluation
of correlation functions in {\it generic} phase of the Hermitean matrix model
in terms of peculiar ``check-operators".
Explicit expressions for the lowest correlation functions
and correlation check-operators are put together in Tables 1 and 2
to demonstrate that the latter ones
(but not the former) look just the same as the Gaussian expressions,
surveyed in \cite{amm1}.
{\bf Explicit formulas for the check-operator multidensities
in Tables 1 and 2
can be considered as the main result of this paper.
Our main hypothesis that the appropriate check-operator
multidensities exist in general, is formulated in s.\ref{hyp}.}
In s.\ref{corf} we discuss definitions
and explicitly specify the {\it ordering} which should be used to
obtain the results
in Tables 1 and 2.
The derivations in their present form are not {\it conceptually}
satisfactory, further work is needed to considerably improve them.
In the last s.\ref{redv} we discuss the variety of solutions to reduced
Virasoro constraints (\ref{redvirF}), i.e. the entire variety of phases,
which admit the genus expansion. Even for $W(z) = \frac{1}{2}z^2$
the variety is non-trivial: it
includes arbitrary linear combinations of the Gaussians.

\section{Correlation functions and check operators \label{corf}}

\subsection{Full and connected correlators}

The main task of quantum field theory in application to particular
model, to the Hermitean matrix model in our case, is to provide
expressions for the correlation functions.
Partition function is a generating function of correlators of a particular
complete set of operators.
The partition function ${\cal Z}(t)$, defined in (\ref{virc}) is
associated in this way with single-trace operators, like
${\rm Tr} \Phi^k$,  or ${\rm Tr}\ e^{ s\Phi}$,
or ${\rm Det}^{\pm 1} (z\cdot I- \Phi)$,
or ${\rm Tr}(z\cdot I- \Phi)^{-1}$.
Correlation functions of such operators can be obtained by taking
derivatives of ${\cal Z}(t)$ w.r.t. its variables $t$; for example,
the insertion of ${\rm Tr} \Phi^k$ corresponds to the action of
$\frac{\partial}{\partial t_k}$,
that of ${\rm Tr}\ e^{s\Phi}$ -- to the action of
$\sum_{k=0}^\infty \frac{s^k}{s!}\frac{\partial}{\partial t_k}$,
that of ${\rm Det}^{\pm 1} (z\cdot I- \Phi)$ -- to the action of
Miwa transform generator
$\exp\left(\mp\sum_{k=0}^\infty
\frac{1}{kz^{k}}\frac{\partial}{\partial t_k}\right)$,
that of ${\rm Tr}(z I- \Phi)^{-1}$ -- to the action of
\be
\hat \nabla(z) = \sum_{k=0}^\infty
\frac{1}{z^{k+1}}\frac{\partial}{\partial t_k}
\label{hatnabla}
\ee
and so on.

In this paper we consider the operators of the last type,
and use the operators $\hat\nabla(z)$, introduced in (\ref{hatnabla})
to generate the ``full" and ``connected" correlation functions,
\be
K_W^{(\cdot|m)}(z_1,\ldots,z_m;g) =
\left.
{\cal Z}_W(t;g)^{-1}\hat\nabla(z_1)\ldots \hat\nabla(z_m){\cal Z}_W(t;g)
\right|_{t=0} =
\sum_{p=0}^\infty
g^{2(p-m)}
K_W^{(p|m)}(z_1,\ldots,z_m)
\label{corK}
\ee
and
\be
\rho_W^{(\cdot|m)} (z_1,\ldots,z_m;g) =
\left.
\hat\nabla(z_1)\ldots \hat\nabla(z_m){\cal F}_W(t;g)
\right|_{t=0} =
\sum_{p=0}^\infty
g^{2p}\rho_W^{(p|m)}(z_1,\ldots,z_m)
\label{corrho}
\ee
where the prepotential ${\cal F}_W(t;g) = g^2\log {\cal Z}_W(t;g) =
g^2\log {\cal Z}(t-T;g)$.

The correlation functions $K_W$ and $\rho_W$ are related by\footnote{
In particular,
$$
K_W^{(p|1)}(z) = \rho_W^{(p|1)}(z),
$$
$$
K_W^{(p|2)}(z_1,z_2) =
\sum_{p_1+p_2 = p}\rho_W^{(p_1|1)}(z_1)\rho_W^{(p_2|1)}(z_2) +
\rho_W^{(p-1|2)}(z_1,z_2),
$$
$$\ldots$$
Note that the genus-$0$ connected double-point correlator
$\rho_W^{(0|2)}(z_1,z_2)$ contributes only to
$K_W^{(1|2)}(z_1,z_2)$, not to $K_W^{(0|2)}(z_1,z_2)$.
To avoid possible confusion, let us emphasize that the index $p$
can be interpreted as genus only in $\rho^{(p|m)}$, but {\it not}
in $K^{(p|m)}$ or $\check K^{(p|m)}$ and $\check\rho^{(p|m)}$
(to be introduced below).

Eq. (\ref{With}) is just a simple generalization of the following relationship:
\be
e^{-g^{-2}F}\p^me^{-g^{-2}F}=\sum_{k=1}^m
\sum_{\stackrel{1\leq m_1<\ldots <m_k}{m = \nu_1m_1+\ldots+\nu_km_k}}
\frac{g^{-2\nu}m!}{\nu_1!(m_1!)^{\nu_1}\ldots \nu_k!(m_k!)^{\nu_k}}\left(\p^{m_1}F\right)^{\nu_1}\dots(\p^{m_k}F)^{\nu_k}
\label{difexp}
\ee
}
\be
K_W^{(\cdot|m)}(z_1,\ldots,z_m;g) =\\=
\sum_\sigma^{m!} \ \sum_{k=1}^{m} \
\sum_{\nu_1,\ldots,\nu_k = 1}^\infty
\sum_{p_1,\ldots,p_\nu=0}^\infty g^{2(p_1+\ldots +p_\nu-\nu)}
\left(\
\sum_{\stackrel{m_1,\ldots,m_k}{m = \nu_1m_1+\ldots+\nu_km_k}}
\frac{1}{\nu_1!(m_1!)^{\nu_1}\ldots \nu_k!(m_k!)^{\nu_k}}
 \right.\times\\ \left.\times
\rho_W^{(p_1|\tilde m_1)}(z_{\sigma(1)},\ldots,z_{\sigma(\tilde m_1)})
\rho_W^{(p_2|\tilde m_2)}
(z_{\sigma(\tilde m_1+1)},\ldots,z_{\sigma(\tilde m_2)})\ldots
\rho_W^{(p_\nu|\tilde m_\nu)}
(z_{\sigma(m-\tilde m_\nu+1)},\ldots,z_{\sigma(m)})
\ \right)
\label{With}
\ee
The sums here are:
over all $m!$ possible permutations $\sigma$ of $m$ points $z_1,\ldots,
z_m$;
over all possible values of $p$-indices of all connected correlators involved;
and over all possible ways to decompose the positive integer $m$
into a sum of ordered integers $0<m_1<m_2<\ldots<m_k$ taken with
non-vanishing multiplicities $\nu_1,\ldots,\nu_k$, i.e.
$m = \nu_1m_1+\ldots+\nu_km_k$;
$\{\tilde m_1, \ldots \tilde m_v\}$ denotes the sequence of
$\nu = \nu_1+\ldots+\nu_k$ elements: $\{m_1,\ldots,m_1;
m_2,\ldots m_2;  \ldots;m_k, \ldots, m_k\}$.
Explicit examples of formula (\ref{With}) for small $m$ can be found in
Table 1.
\footnote{
The relevant decompositions of the first natural numbers are:
$$1 = 1\cdot 1,$$
$$2 = 2\cdot 1 = 1\cdot 2,$$
$$3 = 3\cdot 1 = 1\cdot 1 + 1\cdot 2 = 1\cdot 3,$$
$$4 = 4\cdot 1 = 2\cdot 1 + 1\cdot 2 = 1\cdot 1 + 1\cdot 3 = 1\cdot 4 =
2\cdot 2$$}

The Virasoro constraints provide recurrent relations for the connected
correlators $\rho_W$, which can be explicitly evaluated and then
used to construct the full correlators $K_W$.

\subsection{From correlators $K$ to operators $\check K$}

Our task is to find how the correlation functions depend
not only on their explicit arguments $z$, but also on the $T$-variables,
which enter through $W(z) = \sum_{k=0}^{n+1}T_kz^k$ and also
through additional arbitrary function, the bare prepotential $F(T;g) =
\sum_{p=0}^{\infty} g^{2p} F^{(p)}(T)$ introduced in
(\ref{barpre}) above.
Because of appearance of this arbitrary function, one can at best hope
to represent answers for the correlation functions in the form of
{\it operators}, acting on $Z(T;g) = \exp \left(g^{-2}F(T;g)\right)$,
i.e. express through operators containing derivatives with respect to the
$T$-variables.
Following section 4.3 of ref.\cite{amm1},
we call these operators {\it check operators} and denote
by the ``check" sign to distinguish them from {\it the hat operators}, which
contain $t$-derivatives and are denoted by ``hats". So, the task is to
express the correlation functions defined
in (\ref{corK}) and (\ref{corrho}) with the help of hat-operators
(containing derivatives w.r.t. infinitely many variables $t_k$,
$k= 0,\ldots$)
through the action of check-operators
(containing derivatives w.r.t. only finitely many variables $T_k$,
$k= 0,\ldots, n-1$),
\be
K_W^{(\cdot|m)}(z_1,\ldots,z_m;g) =
\left.
{\cal Z}_W(t;g)^{-1}\hat\nabla(z_1)\ldots \hat\nabla(z_m){\cal Z}_W(t;g)
\right|_{t=0} =
Z(T;g)^{-1} \check K_W^{(\cdot|m)}(z_1,\ldots,z_m) Z(T;g)
\label{corKcheck}
\ee

As explained in \cite{amm1}, from the Virasoro constraints, (\ref{virc})
one can recursively deduce the {\it connected} correlators
$\rho_W^{(p|m)}$ and then (\ref{With}) can be used to evaluate
$K_W^{(\cdot|m)}$. Making use of explicit expressions for
$\rho_W^{(p|m)}$ listed in Table 2, one can obtain explicit expressions
for $K^{(\cdot|1)}$:\footnote{
For the sake of brevity, from now on
we omit the subscript $W$ in $\check R_W(z)$
and $y_W(z)$ when it should not cause a confusion.
Here (see ref.\cite{amm1})
\be
\check R_W(z) = - \sum_{a,b=0} (a+b+2)T_{a+b+2}z^a
\frac{\partial}{\partial T_b}
\ee
and
\be
y_W(z) \equiv \sqrt{W'(z)^2 - 4\left(\check R(z)F^{(0)}\right)}
\ee
In particular, for the Gaussian potential one has
\be
y_G(z) \equiv \sqrt{z^2 - 4\nu}
\ee
}
\be
K_W^{(\cdot|1)}(z;g) = \frac{1}{g^2}\rho_W^{(0|1)}(z) + \rho_W^{(1|1)}(z) +
g^2\rho_W^{(2|1)}(z) + O(g^4) = \nn \\
= \frac{W'(z) - y(z)}{2g^2} - \left(
\frac{y^{\prime\prime}(z)}{4y^2(z)} +
\frac{(\check R y)(z)}{2y^2(z)} - \frac{(\check R F^{(1)})(z)}{y(z)}
\right)
+g^2\left(\frac{(\check R F^{(1)})^2+(\check
R^2F^{(1)})}{y^3}-\right.\\\left.
-3\frac{(\check R y)(\check R F^{(1)})}{y}+
\frac{(\check R F^{(2)})}{y}-\frac{(\check R^2 y)}{y^4}
+\frac{11(\check R y)^2}{4y^5}
+\frac{2y''(\check R y)}{y^5}-
\frac{(\check R y'')}{y^4}-\frac{y'' (\check R F^{(1)})}
{y^4}+\right.\\\left.+
\frac{1}{2y^2}\p^2\left(-\frac{(\check R y )}{2y^2}+
\frac{(\check R F^{(1)})}{y}\right)+
\left(\frac{5}{16}\frac{( y'')^2}{ y^5}-\frac{1}{8
y^2}\p^2\left(\frac{y''}{y^2}\right)-\frac{1}{8}
\frac{ y ^{(4)}}{ y^4}\right)\right)+O(g^4)
\label{KW.1}
\ee
This should now be represented as
\be
K_W^{(\cdot|1)}(z;g) = Z(T;g)^{-1} \check K_W^{(\cdot|1)}(z;g) Z(T;g)
\ee
moreover, according to section 4.3 of ref.\cite{amm1},
$\check K_W^{(\cdot|1)}(z;g)$ does not depend on the bare
prepotential, but only on $T$'s and $\partial/\partial T$.
Again, making use of the explicit formulas for
$Z(T;g)^{-1} \check y(z;g) Z(T;g)$ from section \ref{Hany}
and $\check\rho^{(p|1)}(z;g)$
from Table 2,
one can rewrite (\ref{KW.1}) as
\be
K_W^{(\cdot|1)}(z;g) =
Z(T;g)^{-1} \left( \frac{W'(z) - \check y(z;g)}{2g^2} -
\frac{1}{4\check y^2(z;g)}\check y^{\prime\prime}(z;g) +\right.\\\left.+
g^2\left( \frac{5}{16}\frac{(\check y^{\prime\prime})^2}{\check y^5} -
\frac{1}{8\check y^2}
\partial^2\left(\frac{\check y^{\prime\prime}}{\check y^2}\right) -
\frac{1}{8}\frac{\check y^{\prime\prime\prime\prime}}{\check y^4} \right)
+ O(g^4)\right) Z(T;g) =
Z(T;g)^{-1} \check K_W^{(\cdot|1)}(z;g) Z(T;g)
\label{Krho2}
\ee
Here
\be
\check y(z;g) = \sqrt{W'(z)^2 - 4g^2\check R(z)}
\label{checky}
\ee
In this way, one gets rid of the terms with explicitly present
operators $\check R$ and prepotentials $F^{(p)}$, and significantly
simplify the formulas.

\subsection{From $\check K$ to $\check\rho$: first examples}

The connected correlators $\rho$ are more "fundamental" than the
full $K$. Therefore, it is natural to wonder if one can find
check-operator analogues of $\rho$'s, once we see that
check-operator counterparts of $K$ do exist and can be of some use.

This means that, at the r.h.s. of  (\ref{Krho2}), we would like
to put
\be
\check K_W^{(\cdot|1)}(z;g) =
\sum_{p=0}^\infty g^{2p-2}\check\rho_W^{(p|1)}(z;g)
\label{Krho1}
\ee

In this way, one gets rid of the terms with explicitly present
operators $\check R$ and prepotentials $F^{(p)}$, the relevant
check-operators $\c \rho^{(\cdot|p)}$ are expressed through $\check y(z;g)$
only (with the single exception of $\check\rho_W^{(0|1)}(z;g)$,
which also contains $W'(z)$.) Thus, the check-operator $\c K^{(\cdot|p)}$ is a
polynomial in $W'$ of degree $p$. Instead, the $g$ dependence is now distributed between
explicit factors like $g^{2p-2}$ and an additional $g$-dependence
of $\check y(z;g)$. This, however, allows check-operators
$\check \rho_W^{(p|m)}$ to look exactly the same (modulo ordering)
as the corresponding Gaussian multidensities $\rho_G^{(p|m)}$,
which are all expressed through $y_G$ only.

In this paper we suggest {\it a hypothesis} that
{\bf eq.(\ref{Krho1}) is indeed true in all orders in $g^2$ and, moreover,
similar expansions hold for all $\check K_W^{(\cdot|m)}(z_1,\ldots,z_m;g)$:
they can be all expressed through multilinear combinations
of check operators $\check\rho_W^{(p|m)}$, which (for $(p|m)\neq (0|1)$)
depend only on $\check y(z;g)$ and its $z$-derivatives in exactly the same
way as $\rho_G^{(p|m)}$ depends on $y_G(z)$}.
However, even to formulate this hypothesis, one needs to introduce
some ordering prescription for products of check-operators, which
we denote through $\ :\ \ :\ $ and which is not, as usual, unique.
We distinguish three levels of ordering: (*) the order in which different
$\check\rho$ stand in the products, (**) the order in which $W'$,
and $\check y$ stand and (***) the order among $\check y$ and its
derivatives. Different ordering prescriptions lead to different explicit
formulas for $\check\rho$, and our hypothesis states that there exist
orderings, when these expressions contain $\check y$, it derivative
and nothing else, except for a few $W'$.
Except for a brief comment (see (\ref{wrho02}) and (\ref{wrho12}) below),
we do not discuss
the freedom at level (**), just fix it as in Tables 1 and 2. At level (**)
we request that all $W'$ in a product are carried to the left of all
(derivatives of) $\check y$. If one leaves $W'$ at their places
(we comment on this possibility in the
end of this section), everything
would also work, but in expressions for $\check\rho^{(p|m)}$ some $\check y$
should be substituted by $-2\check\rho^{(1|0)} = \check y - W'$, i.e. $W'$
should appear in explicit expressions for some $\rho^{(p|m)}$. There is nothing
bad in this, we just, somewhat arbitrarily, make the choice to eliminate
these dependencies and, thus, to reduce the freedom at level (**).
As to (*), a possible (though again not unique) option is to define
$\check K^{(\cdot|m)}(z_1,\ldots,z_m|g)$ recursively and put all the
operators containing $z_1$, say, to the left of all others. In this way, we
explicitly break the symmetry of $\check K^{(\cdot|m)}$ and
$\rho^{(p|m)}$ with respect to permutations of its arguments $z_1,\ldots,z_m$.

The ordering makes its first appearance in the
next after (\ref{KW.1}) example,
\be
K_W^{(\cdot|2)}(z_1,z_2;g) =
g^{-4}\rho_W^{(\cdot|1)}(z_1)\rho_W^{(\cdot|1)}(z_2) +
g^{-2}\rho_W^{(\cdot|2)}(z_1,z_2)=\nn\\
=g^{-4}\rho_W^{(0|1)}(z_1)\rho_W^{(0|1)}(z_2) +
g^{-2}\left(\rho_W^{(0|2)}(z_1,z_2) +
\rho_W^{(0|1)}(z_1)\rho_W^{(1|1)}(z_2) +
\rho_W^{(1|1)}(z_1)\rho_W^{(0|1)}(z_2) \right)+O(g^0) = \nn \\ =
\frac{1}{g^4}\frac{W'(z_1)W'(z_2)-W'(z_1) y(z_2)-W'(z_2) y(z_1)+
y(z_1) y(z_2)}{4} + \nn \\
+\frac{1}{g^2}\left(
-\frac{1}{2y(z_1)}\left(\frac{\p}{\p z_2}\frac{y(z_1)-y(z_2)}{z_1-z_2}+\left(\c
R(z_1)y(z_2)\right)\right)+\right.\nn\\
+\left(-\frac{y(z_1)^{\prime\prime}}{4y(z_1)^2} -
\frac{(\check R(z_1) y(z_1))}{2y(z_1)^2} +
\frac{(\check R(z_1) F^{(1)})}{y(z_1)}\right)\frac{W'(z_2) -
y(z_2)}{2}+\nn\\
+\left.\left(-\frac{y(z_2)^{\prime\prime}}{4y(z_2)^2}-
\frac{(\check R(z_2) y(z_2))}{2y(z_2)^2} +
\frac{(\check R(z_2) F^{(1)})}{y(z_2)}\right)\frac{W'(z_1) - y(z_1)}{2}\right)+O(g^0)
=\nn\\
=Z(T;g)^{-1} \check K_W^{(\cdot|2)}(z_1,z_2;g) Z(T;g)
\label{KW.2}
\ee
where
\be
\check K_W^{(\cdot|2)}(z_1,z_2;g) =
g^{-4}:\check\rho_W^{(\cdot|1)}(z_1)\check\rho_W^{(\cdot|1)}(z_2): +
g^{-2}\c \rho_W^{(\cdot|2)}(z_1,z_2) = \nn\\ =
\frac{1}{g^4}\frac{W'(z_1)W'(z_2)-W'(z_1)\c y(z_2)-W'(z_2)\c y(z_1)+\c y(z_1)\c y(z_2)}{4}+
\frac{1}{g^2}\left(-\frac{1}{2\check y(z_1)}\frac{\p}{\p z_2}\frac{\check y(z_1)-\check
y(z_2)}{(z_1-z_2)}\right.+\nn\\
+\left.\left(-\frac{W'(z_1)-\c y(z_1)}{2}\frac{1}{4\c y(z_2)^2}\c y(z_2)''
-\frac{W'(z_2)}{2}\frac{1}{4\c y(z_1)^2}\c y(z_1)''
+\frac{1}{4\c y(z_1)^2}\c y(z_1)\frac{\c y(z_2)}{2}''
\right)\right)+O(g^0)
\ee
so that
\be
\check\rho_W^{(0|2)}(z_1,z_2;g) =  -\frac{1}{2\check y(z_1;g)}
\frac{\partial}{\partial z_2}\frac{\check y(z_1;g)-\check
y(z_2;g)}{z_1-z_2}
\ee
Alternatively, one could consider the natural ordering
\be
\check K_W^{(\cdot|2)}(z_1,z_2;g) =
g^{-4}\check\rho_W^{(\cdot|1)}(z_1)\check\rho_W^{(\cdot|1)}(z_2) +
g^{-2}\rho_W^{(\cdot|2)}(z_1,z_2)
\ee
This would correspond to a different ordering at level (**) and provide
us with the other expressions for $\c\rho^{(p|k)}$,
\be
\check{\tilde\rho}_W^{(0|2)}(z_1,z_2;g) =
\frac{1}{\check y(z_1;g)}
\frac{\partial}{\partial z_2}\frac{\check \rho^{(0|1)}(z_1;g)-\check
\rho^{(0|1)}(z_2;g)}{z_1-z_2}
\label{wrho02}
\ee
and
\be
\check{\tilde\rho}_W^{(1|2)}(z_1,z_2;g) =
\frac{1}{\check y_1}\left[
\left(4\frac{1}{4\c y_1^2}\c y_1'' -\frac{1}{2\check y_1}\p_1^2\right)
(-\check{\tilde\rho}_W^{(0|2)}(z_1,z_2;g))
+\right.\nn\\
+\left.\frac{\p }{\p z_2}\frac{1}{z_1-z_2}\left(\frac{1}{4\c y_2^2}\c y_2''-\frac{1}{4\c y_1^2}\c y_1''+
\frac{1}{y_1}
\left(\check{\tilde\rho}_W^{(0|2)}(z_1,z_1;g)-\check{\tilde\rho}_W^{(0|2)}(z_1,z_2;g)
\right)
\right)\right]
\label{wrho12}
\ee
Similar expressions can be found for higher check operators.

\subsection{From $\check K$ to $\check\rho$: general case \label{REC}}

In principle, when introducing
$\check\rho$-operators, we have different possibilities of definition, preserving
one or another kind of their relation to $\check K$'s.
They could be defined similarly to (\ref{corKcheck}) from (\ref{corrho}),
so that eq.(\ref{neeq}) below becomes an equality.
However, it appears more interesting {\it instead} of
preserving the equations, to require for $\c \rho^{(\cdot|k)}$
to be the same (up to ordering) as the Gaussian functions $\rho_G^{(\cdot|k)}$.
We can construct recursively an operator modification of expression (\ref{With}).
From now on, the normal ordering puts all $W'$ to
the left of all (derivatives of) $\c y$ and $\c R$ which appear in equations.
Since the recurrent equations for $\c K$ are linear (see (\ref{recK})),
they coincide with the equations for
operators $\c K$. The equations for functions $\rho$ are not linear.
Thus, for the operators $\c \rho$ we should choose some ordering on level (*).
In this section, it is convenient to use the inverse ordering of the variables.

We define check operators $\c \rho^{(\cdot|k)}$ with the help of the operation
$J(z_p)[\dots]$,
which transform $n$-point operators into $(n+1)$-point, and is defined as follows:
\be
\label{JK}
J(z_p)[:\c K^{(\cdot|p-1)}(z_{p-1},\dots)\dots:]=:\c
K^{(\cdot|p)}(z_p,z_{p-1},\dots)\dots:\\
J(z_p)[:\c \rho^{(\cdot|p-1)}(z_{p-1},\dots)\dots:]=:(\c B^{(p)}(z_p,z_{p-1},\dots)+\c
\rho^{(\cdot|p)}(z_p,z_{p-1},\dots))\dots:
\ee
Here $\c B^{(p)}$ is the check counterpart of the quadratic in
$\mathcal{F}$ term of the equation (\ref{virder}),
\be
\c B^{(p)}(z_p,z_{p-1},z_{p-2},\dots,z_1)=
\sum_{k=1}^{p-1}\sum_{I\in K_{k-2}}\c \rho^{(\cdot|k)}(z_p,z_{I})\c
\rho^{(\cdot|p-k)}(z_{p-1},z_{K-I})
\ee
We define the operator $\c \rho^{(\cdot|p)}$ by the equation
\be
\check K^{(\cdot|p)}(z_p,\dots,z_1)=\check A^{(p)}(z_p,\dots,z_1)+\check
B^{(p)}(z_p,\dots,z_1)+\check \rho^{(\cdot|p)}(z_p,\dots,z_1)
\ee
where
\be
\check A^{(p)}(z_p,z_{p-1},\dots,z_1)=J(z_p)[\c A^{(p-1)}(z_{p-1},\dots,z_1)+\c
B^{(p-1)}(z_{p-1},\dots,z_1)]\label{Ap}
\ee
According to our definitions, the equation
\be
\c A^{(p)}(z,z,\dots)+\c B^{(p)}(z,z,\dots)+ \c \rho^{(\cdot|p)}(z,z,\dots)=\nn\\
=W'(z) \left(\c A^{(p-1)}(z,\dots)+\c B^{(p-1)}(z,\dots)+\c
\rho^{(\cdot|p-1)}(z,\dots)\right)
+\c R(z) \c K^{(\cdot|p-2)}(\dots)+\\+
\sum_{i=1}^{p-2}\frac{\p}{\p z_i}
\frac{\c K^{(\cdot|p-2)}(z,\dots,\c z_i,\dots)-\c K^{(\cdot|p-2)}(z_i,\dots,\c z_i,\dots)}{z-z_i}
\ee
reduces to the equations for the connected check operators
\be
\check B^{(p+1)}(z,z,z_{p-1}\dots,z_1)+\check \rho^{(\cdot|p+1)}(z,z,z_{p-1}\dots,z_1)=\nn\\
=W'(z)\check
\rho^{(\cdot|p)}(z,z_{p-1},\dots,z_1)+\sum_{i=1}^{p-1}
\frac{\p}{\p z_i}\frac{\check \rho^{(\cdot|p-1)}(z,\dots,\check z_i,\dots,z_1)
-\check \rho^{(\cdot|p-1)}(z_i,\dots,\check z_i,\dots,z_1)}{z-z_i}
\label{eqconcr}
\ee
which, up to ordering, coincide with equation (\ref{virder}). This construction
starts with
\be
\c A^{(2)}=0,\ \ \ \
\c B^{(2)}(z_2,z_1)=:\c\rho^{(\cdot|1)}(z_2)\c\rho^{(\cdot|1)}(z_1):
\ee
that is,
\be
\check K^{(\cdot|2)}(z_2,z_1)=
\check \rho^{(\cdot|2)}(z_2,z_1)+:
\check \rho^{(\cdot|1)}(z_2)\check \rho^{(\cdot|1)}(z_1):
\label{K2}
\ee
This means that the equation
\be
\check K^{(\cdot|2)}(z,z)=W'(z)\check \rho^{(\cdot|1)}(z)-\check R(z)
\label{nc2}
\ee
is already of the form (\ref{eqconcr}).
At the next step of the recursion one has
\be
\c A^{(3)}(z_3,z_2,z_1)=J(z_3)[\c B^{(2)}(z_2,z_1)]=
:\c K^{(\cdot|2)}(z_3,z_2)\c \rho^{(\cdot|1)}(z_1):
\ee
\be
\check B^{(3)}(z_3,z_2,z_1)=:\check\rho^{(1)}(z_3)\check \rho ^{(\cdot|2)}(z_2,z_1):+
:\check\rho^{(\cdot|2)}(z_3,z_1)\check\rho^{(\cdot|1)}(z_2):
\ee
and
\be
\check K^{(\cdot|3)}(z_3,z_2,z_1)=\check\rho^{(\cdot|3)}(z_3,z_2,z_1)+
\c A^{(3)}(z_3,z_2,z_1)+\check B^{(3)}(z_3,z_2,z_1)=\nn\\
=\check\rho^{(\cdot|3)}(z_3,z_2,z_1)
+:\check\rho^{(\cdot|1)}(z_3)\check\rho^{(\cdot|1)}(z_2)\check\rho^{(\cdot|1)}(z_1):+\nn\\
+:\check\rho^{(\cdot|1)}(z_3)\check\rho^{(\cdot|2)}(z_2,z_1):+:\check\rho^{(\cdot|2)}(z_3,z_2)\check\rho^{(\cdot|1)}(z_1):+
:\check\rho^{(\cdot|2)}(z_3,z_1)\check\rho^{(\cdot|1)}(z_2):=\nn\\
=\check\rho^{(\cdot|3)}(z_3,z_2,z_1)+:\check K^{(\cdot|2)}(z_3,z_2)\check\rho^{(\cdot|1)}(z_1):+
:\check\rho^{(\cdot|1)}(z_3)\check \rho ^{(\cdot|2)}(z_2,z_1):+
:\check\rho^{(\cdot|2)}(z_3,z_1)\check\rho^{(\cdot|1)}(z_2):
\label{K3}
\ee

Substituting expressions (\ref{K2}) and (\ref{K3}) into the equation
\be
\check K^{(\cdot|3)}(z,z,z_1)=W'(z)\check K^{(\cdot|2)}(z,z_1)-\check R (z)\check
\rho^{(\cdot|1)}(z_1)+\frac{\p}{\p z_3}\frac{\check \rho^{(\cdot|1)}(z)-\check \rho ^{(\cdot|1)}(z_1) }{z-z_1}
\label{nc3}
\ee
one detects that $\check A^{(3)}(z,z,z_1)$ is canceled by the terms
$W'(z)\check B^{(2)}(z,z_1)$ and $\check
R(z)\rho^{(\cdot|1)}(z_1)$. Here we use the following identity
\be
-\c R(z) \c \rho^{(\cdot|1)}(z_1)+\frac{\p}{\p z_1}\frac{\check
\rho^{(\cdot|1)}(z)-\c\rho^{(\cdot|1)}(z_1)}{z-z_1}=
-:\c R(z) \c \rho^{(\cdot|1)}(z_1):+\frac{\p}{\p z_1}\frac{\check
\rho_s^{(\cdot|1)}(z)-\c\rho_s^{(\cdot|1)}(z_1)}{z-z_1}
\ee
where, by definition,
\be
\c\rho_s^{(\cdot|1)}(z)=\c\rho^{(\cdot|1)}(z)-\frac{W'(z)}{2}
\ee
The equation one gets
\be
\check\rho^{(\cdot|3)}(z,z,z_1)+:\check\rho^{(\cdot|1)}(z)\check\rho^{(\cdot|2)}(z,z_1):+
:\check\rho^{(\cdot|2)}(z,z_1)\check\rho^{(\cdot|1)}(z):
=W'(z)\check
\rho^{(\cdot|2)}(z,z_1)+\frac{\p}{\p z_1}\frac{\check \rho^{(\cdot|1)}_s(z)-\check \rho^{(\cdot|1)}_s(z_1)}{z-z_1}
\ee
coincides with the equation for the Gaussian three-point function.
At the next step one has
\be
\check A^{(4)}(z_4,z_3,z_2,z_1)=
:\check K^{(\cdot|3)}(z_4,z_3,z_2)\check\rho^{(\cdot|1)}(z_1):+
:\check K^{(\cdot|2)}(z_4,z_3)\check \rho ^{(\cdot|2)}(z_2,z_1):+\nn\\
+:\left(\check K^{(\cdot|3)}(z_4,z_3,z_1)-\check
K^{(\cdot|2)}(z_4,z_3)\check\rho^{(\cdot|1)}(z_1)\right)\check\rho^{(\cdot|1)}(z_2):=\nn\\
=:\left(\check K^{(\cdot|3)}(z_4,z_3,z_2)-\check
K^{(\cdot|2)}(z_4,z_3)\check\rho^{(\cdot|1)}(z_2)\right)\check \rho^{(\cdot|1)}(z_1):+\nn\\
+:\left(\check K^{(\cdot|3)}(z_4,z_3,z_1)-\check
K^{(\cdot|2)}(z_4,z_3)\check\rho^{(\cdot|1)}(z_1)\right)\check \rho^{(\cdot|1)}(z_2):+
:\check K^{(\cdot|2)}(z_4,z_3)\check K^{(\cdot|2)}(z_2,z_1):
\ee
and
\be
\c B^{(\cdot|4)}(z_4,z_3,z_2,z_1)=:\check \rho^{(\cdot|2)}(z_4,z_2)\check \rho^{(\cdot|2)}(z_3,z_1):+
:\check \rho^{(\cdot|2)}(z_4,z_1)\check \rho^{(\cdot|2)}(z_3,z_2):+\nn\\
+:\check \rho^{(\cdot|1)}(z_4)\check \rho^{(\cdot|3)}(z_3,z_2,z_1):+
:\check \rho^{(\cdot|3)}(z_4,z_2,z_1)\check \rho^{(\cdot|1)}(z_3):
\ee
Thus,
\be
\check K^{(\cdot|4)}(z_4,z_3,z_2,z_1)=
\check A^{(4)}(z_4,z_3,z_2,z_1)+\check B^{(4)}(z_4,z_3,z_2,z_1)+\check
\rho^{(\cdot|4)}(z_4,z_3,z_2,z_1)=\nn\\
=:\left(\check K^{(\cdot|3)}(z_4,z_3,z_2)-\check
K^{(\cdot|2)}(z_4,z_3)\check\rho^{(\cdot|1)}(z_2)\right)\check \rho^{(\cdot|1)}(z_1):+\nn\\
+:\left(\check K^{(\cdot|3)}(z_4,z_3,z_1)-\check
K^{(\cdot|2)}(z_4,z_3)\check\rho^{(\cdot|1)}(z_1)\right)\check \rho^{(\cdot|1)}(z_2):+\nn\\
+:\check K^{(\cdot|2)}(z_4,z_3)\check K^{(\cdot|2)}(z_2,z_1):+
:\check \rho^{(\cdot|2)}(z_4,z_2)\check \rho^{(\cdot|2)}(z_3,z_1):+
:\check \rho^{(\cdot|2)}(z_4,z_1)\check \rho^{(\cdot|2)}(z_3,z_2):+\nn\\
+:\check \rho^{(\cdot|1)}(z_4)\check \rho^{(\cdot|3)}(z_3,z_2,z_1):+
:\check \rho^{(\cdot|3)}(z_4,z_2,z_1)\check \rho^{(\cdot|1)}(z_3):+\check
\rho^{(\cdot|4)}(z_4,z_3,z_2,z_1)
\label{K4}
\ee

Explicitly, using the definitions (\ref{K2}),  (\ref{K3}), and
formulas (\ref{nc2}), (\ref{nc3}) and  (\ref{K4}), one can bring the equation
\be
\c K^{(\cdot|4)}(z,z,z_2,z_1)=W'(z)\c K^{(\cdot|3)}(z,z_2,z_1)-\c R(z)\c
K^{(\cdot|2)}(z_2,z_1)+\nn\\
+\frac{\p}{\p z_1}\frac{\c K^{(\cdot|2)}(z,z_2)-\c
K^{(\cdot|2)}(z_1,z_2)}{z-z_1}+\frac{\p}{\p z_2}\frac{\c K^{(\cdot|2)}(z,z_1)-\c
K^{(\cdot|2)}(z_2,z_1)}{z-z_2}
\ee
to the equation
\be
\check \rho^{(\cdot|4)}(z,z,z_2,z_1)+:\check \rho^{(\cdot|2)}(z,z_2)\check \rho^{(\cdot|2)}(z,z_1):+
:\check \rho^{(\cdot|2)}(z,z_1)\check \rho^{(\cdot|2)}(z,z_2):+\nn\\
+:\check \rho^{(\cdot|1)}(z)\check \rho^{(\cdot|3)}(z,z_2,z_1):+
:\check \rho^{(\cdot|3)}(z,z_2,z_1)\check \rho^{(\cdot|1)}(z):=\nn\\
=W'(z)\check \rho^{(\cdot|3)}(z,z_2,z_1)+\frac{\p}{\p z_2}\frac{\check \rho^{(\cdot|2)}(z,z_1)-\check
\rho^{(\cdot|2)}(z_2,z_1)}{z-z_2}+\frac{\p}{\p z_1}\frac{\check \rho^{(\cdot|2)}(z,z_2)-\check
\rho^{(\cdot|2)}(z_1,z_2)}{z-z_1}
\ee

\subsection{From $\c K$ and $\c \rho$ back to $K$ and $\rho$}
With our definition
\be
\rho_W^{(p|m)}(z_1,\ldots,z_m) \neq
Z(T;g)^{-1} \check\rho_W^{(p|m)}(z_1,\ldots,z_m;g) Z(T;g)
\label{neeq}
\ee
the thing is that, in variance with the l.h.s. of (\ref{neeq}),
its r.h.s. is still $g$-dependent,
\be
Z(T;g)^{-1} \check\rho_W^{(p|m)}(z_1,\ldots,z_m;g) Z(T;g) =
\sum_{q=0}^\infty g^{2q} \left[
Z(T;g)^{-1} \check\rho_W^{(p|m)}(z_1,\ldots,z_m;g) Z(T;g)
\right]^{(q)} = \nn \\ =
\sum_{q=0}^\infty g^{2q} \left[
\check\rho_W^{(p|m)}\right]^{(q)}(z_1,\ldots,z_m)
\label{[rho]def}
\ee
where the last equality introduces a condensed notation for
sandwiching $\check\rho^{(p|m)}$ between $Z(T;g)^{-1}$ and
$Z(T;g)$. Obviously,
\be
\left[\check A_1\ldots \check A_\nu\right]^{(0)} =
\left[\check A_1\right]^{(0)}\ldots \left[\check A_\nu\right]^{(0)}
\label{fact0}
\ee
but
\be
\left[\check A_1\ldots \check A_\nu\right]^{(0)}  \neq
A_1\ldots A_\nu
\ee
since (in general)
\be
A_i \neq \left[\c A_i\right]^{(0)}
\ee

In particular,
\be
\left[:\check\rho^{(1|1)}(z_1)\check\rho^{(0|1)}(z_2):\right]^{(0)} -
\rho^{(1|1)}(z_1)\rho^{(0|1)}(z_2)
\stackrel{(\ref{chvsrho1}) + (\ref{fact0})}{=}
\left[\check\rho^{(0|1)}(z_1)\right]^{(1)}
\left[\check\rho^{(0|1)}(z_2)\right]^{(0)}
\ee
therefore,
\be
\rho^{(0|2)}(z_1,z_2) =
\left[\check\rho^{(0|2)}(z_1,z_2)\right]^{(0)} +
\left[:\check\rho^{(0|1)}(z_1)\check\rho^{(0|1)}(z_2):\right]^{(1)}
-
\left[\check\rho^{(0|1)}(z_1)\right]^{(1)}
\left[\check\rho^{(0|1)}(z_2)\right]^{(0)} -\\-
\left[\check\rho^{(0|1)}(z_1)\right]^{(0)}
\left[\check\rho^{(0|1)}(z_2)\right]^{(1)}
=
-\frac{1}{2y(z_1)}
\frac{\partial}{\partial z_2}\frac{y(z_1) - y(z_2)}{z_1-z_2}
- \frac{(\c R(z_1)y(z_2))}{2y(z_1)}
\ee

The relation between $\rho$ and $\check\rho$ follows from
(\ref{corKcheck}) and the connection between $\c \rho$ and $\c K$,
(\ref{JK})-(\ref{Ap}). For instance, for $m=1$
\be
K_W^{(\cdot|1)}(z) = \sum_{p=0}^\infty g^{2p-2}\rho_W^{(p|1)}(z) =
Z(T;g)^{-1} \check K_W^{(\cdot|1)}(z) Z(T;g) = \nn \\ =
Z(T;g)^{-1}\left(\sum_{p=0}^\infty g^{2p-2}
 \check\rho_W^{(p|1)}(z;g)\right)Z(T;g) =
\sum_{p,q=0}^\infty g^{2p+2q-2} \left[ \check\rho_W^{(p|1)}\right]^{(q)}(z)
\ee
so that
\be
\rho_W^{(0|1)}(z) = \left[\check\rho_W^{(0|1)}\right]^{(0)}(z), \nn \\
\rho_W^{(1|1)}(z) = \left[\check\rho_W^{(1|1)}\right]^{(0)}(z) +
\left[\check\rho_W^{(0|1)}\right]^{(1)}(z), \nn \\
\ldots \nn \\
\rho_W^{(p|1)}(z) = \sum_{q=0}^p \left[\check\rho_W^{(p-q|1)}\right]^{(q)}(z)
\label{chvsrho1}
\ee
In order to define the relations between $\rho$ and $\check\rho$
for $m>1$, one needs to introduce the notation similar to (\ref{[rho]def}),
\be
Z(T;g)^{-1} \ : \check\rho_W^{(p_1|m_1)}(z_1,\ldots,z_{m_1};g)
\ldots \check\rho_W^{(p_\nu|m_\nu)}(z_{m-m_\nu+1},\ldots,z_{m};g):\
Z(T;g) = \nn \\ =
\sum_{q=0}^\infty g^{2q} \left[
:\check\rho_W^{(p_1|m_1)}\ldots \check\rho_W^{(p_\nu|m_\nu)}:
\right]^{(q)}(z_1,\ldots,z_m)
\label{[rhorho]def}
\ee
The order of arguments $z_1,\ldots,z_m$ is essential here.

Then, for $m=2$
\be
K_W^{(\cdot|2)}(z_1,z_2) =
\sum_{p_1,p_2} g^{2p_1+2p_2-4}
\rho_W^{(p_1|1)}(z_1) \rho_W^{(p_2|1)}(z_2) +
\sum_p g^{2p-2}
\rho_W^{(p|2)}(z_1,z_2) = \nn \\  =
Z(T;g)^{-1} \check K_W^{(\cdot|2)}(z_1,z_2) Z(T;g)
 = \nn \\ =
Z(T;g)^{-1}
\left(
\sum_{p_1,p_2} g^{2p_1+2p_2-4}
:\check\rho_W^{(p_1|1)}(z_1;g) \check\rho_W^{(p_2|1)}(z_2;g):
\ +
\sum_p g^{2p-2}\check\rho_W^{(p|2)}(z_1,z_2;g)
\right)
Z(T;g)
\ee
or
\be
K_W^{(0|2)}(z_1,z_2) = \rho_W^{(0|1)}(z_1)\rho_W^{(0|1)}(z_2) =
\left[ :\check\rho_W^{(0|1)}\check\rho_W^{(0|1)}:\right]^{(0)} (z_1,z_2)
\ee

Thus, the 2-point functions $\rho^{(p|k)}$ and $K^{(p|k)}$ are
connected with the two-point operators
 $\c \rho^{(p|k)}$ and $\c K^{(p|k)}$ by the following equations
\be
K_W^{(p|2)}(z_1,z_2) =
\rho_W^{(p-1|2)}(z_1,z_2) +
\sum_{p_1+p_2 = p} \rho_W^{(p_1|1)}(z_1)\rho_W^{(p_2|1)}(z_2)
= \nn \\ =
\sum_{q=0}^{p}\left(
\left[ \check\rho_W^{(p-q-1|2)}\right]^{(q)} (z_1,z_2)
+ \sum_{p_1+p_2 = p-q}
\left[:\check\rho_W^{(p_1|1)}\check\rho_W^{(p_2|1)}:\right]^{(q)} (z_1,z_2)
\right)
\ee

\be
\rho_W^{(p|2)}(z_1,z_2)
=\sum_{q=0}^{p}\left(
\left[ \check\rho_W^{(p-q-1|2)}\right]^{(q)} (z_1,z_2)
+ \sum_{p_1+p_2 = p-q}
\left[:\check\rho_W^{(p_1|1)}\check\rho_W^{(p_2|1)}:\right]^{(q)} (z_1,z_2)
\right)-\nn\\
-\sum_{p_1+p_2=p} \left(\sum_{q_1=0}^{p_1} \left[\check\rho_W^{(p_1-q_1|1)}\right]^{(q_1)}(z_1)
\sum_{q_2=0}^{p_2} \left[\check\rho_W^{(p_2-q_2|1)}\right]^{(q_2)}(z_2)\right)
\ee

Connection of the n-point correlation functions with
the n-point check operators is rather obvious.

Note that, acting on $Z(T;g) =
\exp \left(\sum_{p=0}^\infty g^{2p-2}F^{(p)}(T)\right)$,
the operator $g^2\hat R(z)$ produces the term $\check R(z)F^{(0)}$,
which is of the zeroth order in $g$ so that
\be
\sum_{q=0}^\infty g^{2q} \left[\check y\right]^{(q)}(z) =
Z(T;g)^{-1}\check y(z;g) Z(T;g) = y + g^2\left(\frac{(\check R y)}{y^2} -
\frac{2(\check R F^{(1)})}{y}  \right)
+ \nn \\ +
g^4\left(-\frac{11}{2}\frac{(\check R y)^2}{y^5}
+2\frac{(\check R^2 y)}{y^4}+6\frac{(\check R y)}{y^4}\check RF^{(1)}-2\frac{1}{y^3}(\check
R^2F^{(1)}+(\c R F^{(1)})^2) -
\frac{2(\check R F^{(2)})}{y} \right)+ O(g^6)
\ee

\section{Evaluation of $\rho^{(p|m)}$}

In this section we remind the iterative procedure for solving the Virasoro constraints
(\ref{vircc}), which gives explicit expressions for the (few lowest)
connected correlators $\rho^{(p|m)}$ presented in the Table 2.

Throughout the text we distinguish between
the quantities, which depend on all the variables $t$ (we denote them
by the {\it calligraphic} letters) and operators with $t$-derivatives
(labeled by {\it hats}), and those depending on $T$-variables (they are
denoted by the {\it ordinary} capital letters) and operators with $T$-derivatives
(labeled by {\it checks}).
When both the $T$ and $t$ variables are present (they enter always in
combinations $T_k+t_k$ so that the $t$- and $T$-derivatives coincide), the
subscript $W$ is used to label the $T$-dependence. When $t$-variables are
not present, the $W$-subscript is sometimes omitted, to avoid overloading
formulas.

The connected correlators (``multidensities")
can be deduced recursively by solving the
Virasoro constraints (\ref{virc}), which can be conveniently
rewritten as
\be
\hat {\cal L}_W(z|t) {\cal Z}_W(t) = 0
\label{vircc}
\ee
where
\be
\hat {\cal L}_W(z|t) \equiv \sum_{m\geq -1}^\infty \frac{1}{z^{m+2}}
\hat {\cal L}_m(T+t) =
(-W'(z)+v'(z))\hat \nabla(z)  + g^2\hat\nabla^2(z) - \hat {\cal R}_{W}(z|t)
\ee
potentials
$W(z) = \sum_{k=0}^{n+1} T_kz^k$,
$v(z) = \sum_{k=0}^\infty t_kz^k$
and
\be\label{cR}
\hat {\cal R}_{W}(z|t) =
\sum_{a,b\geq 0}^{\infty}
(a+b+2)(-T_{a+b+2}+t_{a+b+2}) z^a\frac{\partial}
{\partial t_b} = \hat R_{v}(z|t)  - \check R_W(z)
\ee
The last equality holds because the derivatives of the partition function
with respect to $t_k$ and $T_k$ for $k=0,\ldots,n-1$ coincide.
\be
\check R_W(z)= \sum_{a,b\geq 0}^{n-1}
(a+b+2) T_{a+b+2} z^a\frac{\partial}{\partial T_b}
\label{Rop}
\ee
is a {\it check-operator}. Below we often denote it by $\check R(z)$,
omitting the subscript $W$.

The standard way of derivation of the connected correlators $\rho^{(p|m)}$
is surveyed in \cite{amm1} and consists of acting with a number
of operators $\hat\nabla(z_i)$ on (\ref{vircc}) and putting all
$t=0$ afterwards. This provides iterative relations for $\rho^{(p|m)}$,
expressing them through the action of the {\it check}-operators on $\rho^{(p|m)}$
with lower values of $p+m$.

The only relation needed in the process, is
\be
\left[\hat\nabla(z_1), \hat l(z)\right] =
\frac{\partial}{\partial z_1}\left(\frac{\hat\nabla(z) -
\hat\nabla(z_1)}{z-z_1}\right)
\label{[nablal]}
\ee
where
\be
\hat L(z) = \hat l(z) + g^2\hat\nabla^2(z)
\ee
and the ``linear Virasoro operator" is
\be
\hat l(z) = (-W'(z)+v'(z))\hat\nabla(z) - \hat R_{-W+v}(z)
\ee

Rewriting (\ref{vircc}) in terms of the prepotential as
\be
\hat l(z) {\cal F} = \left(\hat\nabla(z){\cal F}\right)^2
+ g^2\hat\nabla^2(z){\cal F}
\label{virccF}
\ee
one directly obtains with the help of (\ref{[nablal]})
\be
\left.\left(W'(z) - (\hat\nabla(z){\cal F}^{(0)})\right)
(\hat\nabla(z){\cal F}^{(0)})\right|_{t=0}
= (\check R(z) {F}^{(0)})
\ee
for the $p=0$, $m=1$ contribution,
where $\check R(z)$ is already the {\it check}-operator defined in
(\ref{Rop}) so that
\be
\rho^{(0|1)}(z) = \left.\hat\nabla(z){\cal F}^{(0)}\right|_{t=0} =
\frac{1}{2}\left(W'(z) -
\sqrt{W'(z)^2 - 4 (\check R(z) {F}^{(0)})}\right) =
\frac{1}{2}\left(W'(z) - y(z)\right)
\ee
and, for all the rest,
\be
\left.\left(W'(z) - 2(\hat\nabla(z){\cal F})
\right)\hat\nabla(z)\hat\nabla(z_1)\ldots\hat\nabla(z_m){\cal F}
\right|_{t=0} =
\nn \\ =
\left.
\check R(z)\hat\nabla(z)\hat\nabla(z_1)\ldots\hat\nabla(z_m){\cal F}
\right|_{t=0} +
\left.
\sum_{\stackrel{I,J \neq 0}{I+J = \{z_1,\ldots z_m\}}}
\left(\hat\nabla(z)\prod_{i\in I}\hat\nabla(z_i){\cal F}\right)
\left(\hat\nabla(z)\prod_{j\in J}\hat\nabla(z_j){\cal F}\right)
\right|_{t=0} + \nn \\ +
\sum_{i=1}^m \frac{\partial}{\partial z_i}\frac{1}{z-z_i}
\left.\left(
\hat\nabla(z)\prod_{j\neq i}\hat\nabla(z_j){\cal F} -
\hat\nabla(z_1)\ldots\hat\nabla(z_m){\cal F}
\right)
\right|_{t=0}
\left.
+ g^2\hat\nabla^2(z)\hat\nabla(z_1)\ldots\hat\nabla(z_m){\cal F}
\right|_{t=0}
\label{virder}
\ee
Substituting now the genus expansion (\ref{gexpF}) for the prepotential
and picking up the contributions of the order  $g^{2p}$ to (\ref{virder}),
one obtains the necessary recurrent relations, which provide the
expressions listed in the Tables.

Similarly to (\ref{virder}), for the full correlators the shifted
Virasoro constraint (\ref{vircc}) implies that
\be
W'(z)K^{(\cdot|m+1)}(z,z_1,\ldots,z_m) -
\check R(z)K^{(\cdot|m)}(z_1,\ldots,z_m) +\\+
\sum_{i=1}^m\frac{\partial}{\partial z_i}
\frac{K^{(\cdot|m)}(z,z_1,\ldots,\check z_i,\ldots, z_m) -
K^{(\cdot|m)}(z_1,\ldots,z_m)}{z-z_i} =
g^2K^{(\cdot|m+2)}(z,z,z_1,\ldots,z_m)
\label{recK}
\ee

\section{Handling $\check y$
\label{Hany}}

\subsection{Algebra generated by $\partial^k W(z)$ and $\partial^l\check
R(z)$}

From the definitions of
\be
W'(z) = \sum_{k=1}^{n+1} kT_kz^{k-1}
\ee
and
\be
\check R(x) = \check R_W(x) = -\sum_{a,b=0} (a+b+2)T_{a+b+2}
x^a \frac{\partial}{\partial T_b}
\ee
one immediately deduce
\be
\left[ \check R(x), W'(z) \right] = -\frac{\partial}{\partial z}
\frac{W'(x)-W'(z)}{x-z}
\label{coreRW}
\ee
and
\be
\left[ \check R(x), \check R(z) \right] =
\left(\frac{\partial}{\partial x}
-\frac{\partial}{\partial z}\right)
\frac{\check R(x)-\check R(z)}{x-z}
\label{coreRR}
\ee
Of course, also
\be
\left[ \check W'(x), \check W'(z) \right] = 0
\label{coreWW}
\ee
It, therefore, follows that
\be
\left[ \check y^2(x;g), \check y^2(z;g) \right] = -4g^2
\left(\frac{\partial}{\partial x}
-\frac{\partial}{\partial z}\right)
\frac{\check y^2(x;g)-\check y^2(z;g)}{x-z}
\label{coreyy2}
\ee
and one observes that $\c y^2$ is nothing but the positive (nilpotent) part of the Virasoro
algebra, while $W'$ is the positive part of the $U(1)$-current\footnote{It can be also
evident from manifest formulas (\ref{cR}) if one defines the $U(1)$-current as
$$
\p\c\phi(z)=\sum kT_kz^{k-1}+\sum z^{-k-1}{\p\over\p T_k}
$$
Then, $\c W'(z)=\left[\p\c\phi (z)\right]_+$ and
$\c y^2(z)=\left[:\left(\p\c\phi(z)\right)^2:\right]_+\equiv
\left[\sum_k{{\c\mathcal{N}}_{k}\over z^{k+2}}\right]_+$ formally truncated to the
finite number of $T_k$'s .
}.
Therefore, the modes of operators $\c y^2$ and $\c R$, say,
\be
{\c\mathcal{N}}_{-k-1}(x):=\frac{-\p^k \c y^2(x)}{4g^2k!}
\ee
form a (nilpotent) subalgebra of the Virasoro algebra,
\be
\left[{\c\mathcal{N}}_{k}(x),{\c\mathcal{N}}_{l}(x)\right]=
(k-l){\c\mathcal{N}}_{k+l}(x),\ \ \ \ k,l\le -2
\ee
In particular, with the help of the general identity
(valid for any $A(x)$, function or
operator)
\be
\left. \partial_x^l \partial_z^{k-1} \frac{A(z) - A(x)}{z-x}\right|_{x=z} =
\frac{l!(k-1)!}{(l+k)!}\partial^{l+k}A(z)
\ee
one can rewrite the loop algebra (\ref{coreRW}) - (\ref{coreWW}) in terms of
the Virasoro harmonics, that is,
derivatives of $\check R(z)$ and $W'(z)$ at a single point $z$,
\be
\left[ \partial^l\check R(z), \partial^k W(z)\right] =
-\frac{l!k!}{(l+k+1)!}\partial^{l+k+2}W(z),\ \ l\geq 0,\ k\geq 1 \nn\\
\left[ \partial^l\check R(z), \partial^k \check R(z)\right] =
\frac{(k-l)l!k!}{(l+k+2)!}\partial^{l+k+2}\check R(z),\ \ l,k \geq 0
\ee

The commutator
\be
\left[ \check y(x;g), \check y(z;g) \right] =
\frac{-g^2}{\c y(z_1)\c y(z_2)}\left(\frac{\partial}{\partial x}
-\frac{\partial}{\partial z}\right)
\frac{\check y^2(x;g)-\check y^2(z;g)}{x-z}+O(g^4)
\label{coreyy}
\ee
describes {\bf an important Lie algebra} which
is related to (\ref{coreyy2}) through
\be
\left[ \check y^2(x;g), \check y^2(z;g) \right] =
\left\{\left\{\check y(x;g),\ \left[ \check y(x;g),
\check y(z;g) \right]
\right\}_+, \check y(z;g) \right\}
\ee
but {\bf a closed expression for (\ref{coreyy}) remains unavailable}
(the first terms of the expansion in $g^2$ written in (\ref{coreyy})
are obtained with the help of eq.(\ref{yche}) below).

It is useful to keep in mind that $\check R$ is just a linear
differential operator satisfying the Leibnitz rule; thus, for example,
\be
(\check R^p y^2)(z) = \sum_{q=0}^p \frac{p!}{q!(p-q)!}
(\check R^q y)(\check R^{p-q} y)(z)
\ee
and
\be
(\check R^p {W'}^2)(z) = \sum_{q=0}^p \frac{p!}{q!(p-q)!}
(\check R^q W')(\check R^{p-q} W')(z) =
(-)^p \sum_{q=0}^p \frac{p!}{2^p (q!(p-q)!)^2}
(\partial^{2q+1}W)(\partial^{2p-2q+1}W)(z)
\ee
since, as a corollary of (\ref{coreRW}),
\be
\left(\check R\ \partial^kW\right)(z) =
\left[\check R(z), \partial^kW(z)\right] =
-\frac{1}{k+1}\partial^{k+2}W(z)
\ee
and
\be
(\check R^q W')(z) = \frac{(-)^q}{(2q)!!}\partial^{2q+1}W(z) =
\frac{(-)^q}{2^q q!}\partial^{2q+1}W(z)
\ee

\subsection{Integral representation for powers of $\check y$}

Operator $\check y(z;g)$ is defined in (\ref{checky}) as
\be
\check y^2(z;g) = W'(z)^2 - 4g^2\check R(z)
\label{checky2}
\ee
Powers of $\check y$ and their action on $Z(T;g)$ can be evaluated
with the help of the integral representation
\be
\check y^{-k} = \frac{1}{\Gamma(k/2)}
\int \frac{ds}{s}s^{k/2} e^{-s\check y^2}
\label{ire}
\ee

\subsection{Toy example of further calculations}

Since, for every given $z$, the check-operator $\check R(z)$ is a linear
differential operator in $T$, and $W'(z)^2$ is a function of $T$, one
can take as a toy example of $\check y^2(z;g)$ just
$w(q) - \hbar\frac{\partial}{\partial q}$. In this toy example
\be
e^{-s\left(w(q) - \hbar\frac{\partial}{\partial q}\right)} =
\exp \left(-\frac{1}{\hbar}\int_{q}^{q+s\hbar} w(x)dx\right)
e^{s\hbar\frac{\partial}{\partial q}}
\ee
The integral in the first factor at the r.h.s. can also be rewritten as
\be
\frac{1}{\hbar}\int_{q}^{q+s\hbar} w(x)dx = \frac{e^{s\hbar\frac{\partial}{\partial q}}-1}
{\hbar\frac{\partial}{\partial q}} w(q)
\ee
The shift operator acts on the toy partition function
$Z(q) = e^{\frac{1}{\hbar}F(q)}$ as follows
\be
Z^{-1}(q) e^{s\hbar\frac{\partial}{\partial q}} Z(q) = Z^{-1}(q)Z(q+s\hbar)
= \exp \frac{1}{\hbar}(F(q+s\hbar) - F(q)) =
\exp \left(\left(e^{s\hbar\frac{\partial}{\partial q}}-1\right)
\frac{1}{\hbar}F(q)\right)
\ee
Thus,
\be
Z^{-1}(q) e^{-s(w(q) - \hbar\frac{\partial}{\partial q})} Z(q) =
\exp \left(\frac{1-e^{s\hbar\frac{\partial}{\partial q}}}
{\hbar\frac{\partial}{\partial q}} \left(w(q) - \frac{\partial F(q)}
{\partial q}\right)\right)
\ee
Not surprisingly, the formulas for $e^{-s\check y^2}$ in the following three
subsections will just reproduce the above expressions of the toy example.

\subsection{Entangling exponential}

Exponential of $\check y^2$ can be handled with the help of the
Campbell-Hausdorff formula \cite{CH},
\be
e^{A}e^{B} =
\exp \left( \int_{0}^1 ds
\frac{\log\left(e^{s\cdot ad_A} e^{s\cdot ad_B}\right)}
{e^{s\cdot ad_A} e^{s\cdot ad_B} - 1}e^{s\cdot ad_A}
\left( A + B\right)\right)
\label{CHf}
\ee
The operators $ad_A$ and $ad_B$ are defined to act as commutators:
$ad_A = \left[A,\ \right]$ and $ad_B = \left[B,\ \right]$.
In our case, $A = -sW'(z)^2$ and $B = 4s g^2\check R(z)$ and
only $ad_B$ acts non-trivially on $[A,B]$: this significantly
simplifies the formula (of which we, actually, need the inverse):
\be
e^{-s\check y^2(z)} =
e^{-s\left(W'(z)^2 - 4g^2\check R(z)\right)} = \nn \\ =
\exp \left(-sW'(z)^2
- \frac{4g^2 s^2}{2!}(ad_{\c R}{W'}^2)(z)
- \frac{(4g^2)^2 s^3}{3!} (ad_{\c R}^2 {W'}^2)(z)
- \ldots \right)
e^{4sg^2 \check R(z)}
= \nn \\ =
\exp \left( -\sum_{p=0}^\infty \frac{(4g^2)^p s^{p+1}}{(p+1)!}
(ad_{\c R}^p {W'}^2)(z)\right)
e^{4sg^2 \check R(z)} =
\exp \left(\frac{1 - e^{4sg^2 ad_{\c R}(z)}}
{4g^2 ad_{\c R}(z)}{W'}^2(z)\right) e^{4sg^2 \check R(z)}
\label{expsum}
\ee
To avoid a confusion, note that here $(ad_{\c R}^p {W'}^2)(z) =
(ad_{\c R}^p(z) {W'(z)}^2)$, i.e. the argument $z$ is the same in
all $\check R(z)$ and $W'(z)^2$.

From (\ref{ire}) and (\ref{expsum}), one gets
\be
\check y^{-k} = \frac{1}{\Gamma(k/2)}
\int \frac{ds}{s}s^{k/2}
\exp \left( -\sum_{p=0}^\infty \frac{(4g^2)^p s^{p+1}}{(p+1)!}
(ad_{\c R}^p {W'}^2)(z)\right) e^{4sg^2 \check R(z)} = \nn \\
= \frac{1}{W'^k} +  4g^2 \left(
\frac{\Gamma(\frac{k}{2}+1)}{\Gamma(\frac{k}{2})W'^{(k+2)}}\c R-
\frac{\Gamma(\frac{k}{2}+2)}{\Gamma(\frac{k}{2})W'^{(k+4)}}\frac{(ad_{\c R} W'^2)}{2}
\right)+\nn\\
+(4g^2)^2 \left(\frac{\Gamma(\frac{k}{2}+4)}{\Gamma(\frac{k}{2})W'^{(k+8)}}\frac{(ad_{\c R}W'^2)^2}{2!2!2!}
-\frac{\Gamma(\frac{k}{2}+3)}{\Gamma(\frac{k}{2})W'^{(k+6)}}\left(\frac{(ad_{\c R}^2
W'^2)}{3!}+
\frac{(ad_{\c R}W'^2)}{2!}\c R\right)+
\right.\nn \\
\left.
\frac{\Gamma(\frac{k}{2}+2)}{\Gamma(\frac{k}{2})W'^{(k+4)}}\frac{\c R^2}{2}
\right) + O(g^6)
\label{yche}
\ee

\subsection{Action of $\check R$-shift operator}

Further, similarly to (\ref{With}) and (\ref{difexp}),
\be
Z^{-1}(T;g) e^{4sg^2 \check R(z)} Z(T;g) = Z^{-1}(T;g)
\sum_{m=0}^\infty \frac{\left(4sg^2\check R\right)^m(z)}{m!} Z(T;g) =
\nn \\ =
\sum_{m = \sum_{i=1}^k \nu_im_i}
\prod_{i=1}^k
\frac{\left((4sg^2)^{m_i}(ad_{\c R}^{m_i}F)(z;g)\right)^{\nu_i}}
{\nu_i!(m_i!)^{\nu_i}} =
\prod_{l=1}^\infty \left(\sum_{\nu_l=0}^\infty
\frac{\left((4sg^2)^{l}(ad_{\c R}^{l}F)(z;g)\right)^{\nu_l}}
{\nu_l! (l!)^{\nu_l}}\right) =\\=
\exp \left(
\sum_{l=1}^\infty \frac{(4g^2s)^l(ad_{\c R}^{l}F)(z;g)}{l!}
\right) =
\exp \left(\left(e^{4sg^2ad_{\c R}(z)} - 1\right)F\right)
= \nn \\ =
\exp \left(4s (ad_{\c R}(z)F^{(0)}) +
g^2\left(4s (ad_{\c R}(z) F^{(1)}) + 8s^2 ad_{\c R}^2(z)F^{(0)}\right)
+ O(g^4) \right)
\label{Ract}
\ee
Note that $F(T;g) = \sum_p g^{2p}F^{(p)}(T)$
here depends on $g$ and includes contributions from all genera.
Powers of operator $\check R$ can be written in various ways:
$(\check R^p F)(z) =
\left[\check R(z),\left[\check R(z),\left[ \ldots [\check R(z),F]\ldots
\right]\right]\right] = ad_{\check R(z)}^p F$.

\subsection{Action of $\check y$ and its powers}

Multiplying (\ref{expsum}) and (\ref{Ract}), one gets, substituting
$l=p+1$,
\be
\exp \left( -\sum_{p=0}^\infty \frac{(4g^2)^p s^{p+1}}{(p+1)!}
(ad_{\c R}^p {W'}^2)(z)\right)
\exp \left(
\sum_{l=1}^\infty \frac{(4g^2s)^l(ad_{\c R}^{l}F)(z;g)}{l!}
\right) = \nn \\ =
\exp \left( -\sum_{p=0}^\infty
\frac{(4g^2)^p s^{p+1}}{(p+1)!} \left(ad_{\c R}^p Y^2(z;g)\right)\right)
= \exp \left(\frac{1-e^{4sg^2ad_{\c R}}}{4g^2ad_{\c R}}Y^2(z;g)\right)
\label{Y}
\ee
where
\be
Y^2(z;g) = W'(z)^2 - 4g^2 (ad_{\c R}F)(z;g) =
y^2(z) - 4\sum_{p=1}^\infty g^{2p} \left(ad_{\c R}(z) F^{(p)}\right)
\ee
and, finally,
\be
Z^{-1}(T;g) \check y^{-k}(z;g) Z(T;g) =
\frac{1}{\Gamma(k/2)} \int \frac{ds}{s}s^{k/2}
\exp \left( -\sum_{p=0}^\infty
\frac{(4g^2)^p s^{p+1}}{(p+1)!} \left(ad_{\c R}^p Y^2\right)(z;g)\right)
\label{ykZ}
\ee

Part of the above calculation can also be applied for evaluating the
action on $Z(T;g)$ of {\it combinations} made from $\check y(z;g)$
and its $z$-derivatives, which enter the expressions for
$\check\rho^{(p|m)}$. Technique, however, remains undeveloped.

Another possibility is to single out the genus zero contribution
\be
Z_0=\exp{\left(\frac{F^{(0)}}{g^2}\right)}
\ee
from the function $Z(T;g)$. Then, from (\ref{Y}) one gets
\be
Z_0e^{-s\c y^2}Z_0=\exp{\left(
-\sum_{p=0}^\infty \frac{(4g^2)^p s^{p+1}}{(p+1)!}
(ad_{\check R}^p y^2)\right)}\exp(-4g^2s\c R)
=e^{-sy^2}\left(1-4g^2\left(\frac{s^2}{2!}ad_{\c R}y^2+s\c R\right)+\right.\nn\\
+\left.(4g^2)^2
\left(\frac{s^3}{2!}(ad_{\c R}y^2) \c R+\frac{s^4}{2!(2!)^2}(ad_{\c R}y^2)^2-
\frac{s^3}{3!}(ad_{\c R}^2y^2)+\frac{s^2}{2!}\c R^2\right)+O(g^6)\right)
\ee
With the help of
\be
\frac{1}{\Gamma(k/2)}
\int \frac{ds}{s}s^{k/2}e^{-sy^2}s^m=
\frac{\Gamma\left(\frac{k}{2}+m\right)}{\Gamma\left(\frac{k}{2}\right)}y^{-(k+2m)}
\ee
one gets
\be
Z_0\c y^{-k}Z_0=
y^{-k}-4g^2\left(\frac{\Gamma\left(\frac{k}{2}+2\right)}{2!\Gamma\left(\frac{k}{2}\right)}y^{-(k+4)}ad_{\c R}y^2
+\frac{\Gamma\left(\frac{k}{2}+1\right)}{\Gamma\left(\frac{k}{2}\right)}y^{-(k+2)}\c R\right)+\nn\\
+(4g^2)^2
\left(
\frac{\Gamma\left(\frac{k}{2}+4\right)}{2!(2!)^2\Gamma\left(\frac{k}{2}\right)}y^{-(k+8)}(ad_{\c R}y^2)^2+
\frac{\Gamma\left(\frac{k}{2}+3\right)}{\Gamma\left(\frac{k}{2}\right)}y^{-(k+6)}
\left(\frac{(ad_{\c R}y^2)}{2!}\c R-\frac{(ad_{\c R}^2y^2)}{3!}\right)+\right.\nn\\
+\left. \frac{\Gamma\left(\frac{k}{2}+2\right)}{2!\Gamma\left(\frac{k}{2}\right)}y^{-(k+4)}\c R^2\right)+O(g^6)
\ee

One can further transform (\ref{ykZ}) with the help of Shur polynomials,
\be
\exp \left(\sum_{p=1}^\infty r_{p} s^{p}\right) =
\sum_{l=0}^\infty s^l {\cal S}_l(r)
\ee
however, in our case,
\be
r_{p+1}(z;g) =
-\frac{(4g^2)^p}{(p+1)!} \left(ad_{\c R}^p(z) Y^2\right)(z;g)
\ee
still depends on $g^2$ in a complicated way (because $Y^2$ is
$g$-dependent), and this formalism is not immediately useful for
handling the $g^2$-expansions.

Since in this paper we need just the first terms of this expansion,
it is simpler to read them directly from (\ref{ykZ}). For example,
\be
Z^{-1}(T;g) \check y(z;g) Z(T;g) =
y + g^2\left(\frac{(ad_{\c R} y)}{y^2} -
\frac{2(ad_{\c R}F^{(1)})}{y}  \right)
- \nn \\ -
g^4\left(\frac{11}{2}\frac{(ad_{\c R}y)^2}{y^5}
-2\frac{(ad_{\c R}^2 y)}{y^4}-6\frac{(ad_{\c R}y)}{y^4}ad_{\c R}F^{(1)}+
\frac{2}{y^3}\left[ad_{\c R}^2F^{(1)}+(ad_{\c R}F^{(1)})^2\right] +
\frac{2(ad_{\c R} F^{(2)})}{y} \right)+ O(g^6)
\ee

\section{Freedom in solving reduced Virasoro constraints \label{redv}}
\subsection{Independent variables}

As a corollary of the shifted and reduced Virasoro constraints (\ref{redvirF}),
the partition function can be represented as
\be
\left.Z(T)\right|_{t=0} = \int dk z(k;\eta_2,\ldots,\eta_n;\hbar)
e^{\frac{1}{\hbar}(kx-k^2w)}
\label{etaparam}
\ee
with an arbitrary function $z$ of $n$ arguments $(k,\eta_2,\ldots,\eta_n)$
and $\hbar$.
Here the $\hat {\cal L}_{-1}$-invariant variables are used,
\be
w = \frac{1}{n+1}\log T_{n+1}, \ \ \ \
\eta_k = T_{n+1}^{-\frac{nk}{n+1}} \left( T_n^k + \ldots \right),\ \ \ \
x  = T_0 + \ldots \sim \eta_{n+1}
\ee
In particular,
\be
\eta_2  = \left(T_n^2 -
\frac{2(n+1)}{n}T_{n-1}T_{n+1}\right)T_{n+1}^{-\frac{2n}{n+1}}, \nn \\
\eta_3  = \left(T_n^3 -
\frac{3(n+1)}{n}T_{n-1}T_nT_{n+1} + \frac{3(n+1)^2}{n(n-1)}T_{n-2}T_{n+1}^2
\right)T_{n+1}^{-\frac{3n}{n+1}}, \nn \\
\eta_4  = \left(T_n^4 -
\frac{4(n+1)}{n}T_{n-1}T_n^2T_{n+1} +
\frac{8(n+1)^2}{n(n-1)}T_{n-2}T_nT_{n+1}^2 -
\frac{8(n+1)^3}{n(n-1)(n-2)}T_{n-3}T_{n+1}^3
\right)T_{n+1}^{-\frac{4n}{n+1}}, \nn \\
\ldots \nn \\
\eta_k = \left(T_n^k + \frac{k(k-2)!}{n!}\sum_{l=1}^{k-1} (-)^l
\frac{(n+1)^l (n-l)!}{(k-l-1)!}
T_{n-l}T_n^{k-l-1}T_{n+1}^l
\right)T_{n+1}^{-\frac{kn}{n+1}}
\ee
The variable $x$ is obtained from $\eta_{n+1}$ by normalization.
The $\hat {\cal L}_{0}$-constraint links the $x$- and $w$-dependencies in
(\ref{etaparam}). Also, $W(\alpha_i)$
is $\hat {\cal L}_{-1}$- and $\hat {\cal L}_{0}$-invariant
for any root $\alpha_i$ of $W'(z)$ in eq.(\ref{rootsW'}).

If
\be
S = \frac{\partial{\cal F}}{\partial T_0} = const
\ee
i.e. is independent of $T_0,\ldots,T_{n+1}$ and $g$,
then,
\be
z(k;\eta_2,\ldots,\eta_n;\hbar) = \delta(k-\hbar S)H(\eta_2,\ldots,\eta_n;\hbar,S)
\ee
and
\be
{\cal F} = \log Z = Sx + \frac{S^2}{n+1}\log T_{n+1} + \log
H(\eta_2,\ldots,\eta_n)
\ee
where $H$ is an arbitrary function of $n-1$ variables (it may depend on $S$ as
well).
A sophisticated counterpart of the Fourier transform with the help of DV partition
functions \cite{DV,DVfollowup} converts $H(\eta_2,\ldots,\eta_n)$ into
an arbitrary function of the peculiar $S_i$-variables, see \cite{amm1}.

\subsection{Changing power of $W$: $n+1=3$ example}

\be
x = T_0 - \frac{9T_1T_2T_3 - 2T_2^3}{27T_3^2} = T_0 +
\frac{2}{27}\frac{T_2^3}{T_3^2}\left(1 -
\frac{9}{2}\frac{T_1T_3}{T_2^2}\right)\\
\eta_2 = \frac{T_2^2 - 3T_1T_3}{T_3^{4/3}} =
\left(\frac{T_2^3}{T_3^2}\right)^{2/3}\left(1 - 3\frac{T_1T_3}{T_2^2}\right)
\ee
As $T_3 \rightarrow 0$, $x$ and $\eta_2$ are both singular, while
\be
x^{(3)} - \frac{2}{27}\eta_2^{3/2} \rightarrow x^{(2)}
\ee
and
\be
Sx^{(3)} + \frac{1}{3}S^2\log T_3 + g(\eta_2) \rightarrow Sx^{(2)} +
\frac{1}{2}S^2\log T_2
\ee
provided
\be
g(\eta_2) \sim -\frac{2}{27}S\eta_2^{3/2} - \frac{1}{4}S^2\log \eta_2 +
O(\eta_2^{-1})
\ee

\subsection{Gaussian $n=1$ case, eq.(\ref{Gprep})}

Here we want to stress that, along with standard solution for the Gaussian
potential
($n+1 = 2$),
\be
\exp \left(g^{-2}F_G(T_0;g)\right) =
\frac{(g/T_2)^{-N^2/2}}{{\rm Vol}(SU(N))}
e^{-NT_0/g}
\label{gaussc}
\ee
there are non-conventional solutions. Namely, any linear combination of
solutions of the form (\ref{gaussc})
\be
Z = \sum_N c_N Z_N, \ \ \ \
Z_N = T_2^{-N^2/2} e^{-NT_0/g}
\ee
where coefficients $c_N$ may depend on $N$, and $N$ is an arbitrary (not
obligatory natural) number.
A particular solution of this form is the $\theta$-function
\be
Z = \theta(T_0|\log T_2)
\ee
For such solutions, $\rho$'s turn into the check-operators, e.g.,
\be
\rho^{(0|1)}(z) = \frac{1}{2Z}\left(2T_2z -
\sqrt{(2T_2z)^2 - 8T_2g^2\partial/\partial T_0}\right) Z
\ee

\section{Summary: the Hypothesis
\label{hyp}}

To summarize, we suggest to consider the following 3-level {\it hypothesis}.

$\bullet$ $F$-independent check-operators $\check
K_W^{(\cdot|m)}(z_1,\ldots,z_m;g)$, satisfying (\ref{corKcheck}),
do exist (this is actually proved in s.4.3 of \cite{amm1}) and can be
expressed only through, somehow ordered, $\check y_W$, its derivatives and
$W'$. The dependence on $W'$ is rather simple: $\check
K_W^{(\cdot|m)}(z_1,\ldots,z_m;g)$ is a polynomial in $W'$ of degree $m$.

$\bullet$ Check-operators $\check\rho_W^{(p|m)}(z_1,\ldots,z_m;g)$ do exist,
related to $\check K$ by the operator identity (\ref{With}),
which is defined recurrently in section \ref{REC}
with some ordering, and $\check\rho_W^{(p|m)}$ are expressed through
$\check y_W$ in exactly the same way (modulo ordering) as the corresponding
Gaussian $\rho_G^{(p|m)}$ are expressed through $y_G$.

$\bullet$ All above-mentioned orderings do not need to be uniquely
defined: there can be orderings at level (*), different from
((\ref{JK})-(\ref{Ap})), but the
change of $\ :\ :\ $ can be
{\it often} compensated by reordering of $\check y$ and their
derivatives inside $\check\rho$'s or, at worst, by substituting some
of $\check y$
by $-2\check\rho^{(0|1)}$. Tables 1 and 2 assume
one particular, but perhaps in no way distinguished, ordering.

The status of this hypothesis at every level is yet unclear; in this paper we
only presented some evidence for existence of a few lowest $\check\rho^{(p|m)}$,
but general proofs (and even convincing arguments) are still lacking.
Without them, one can hardly speak about satisfactory understanding of
non-Gaussian phases; thus, we insist the problem deserves an attention and
further investigations.

\bigskip

\section{Acknowledgements}

Our work is partly supported by Federal Program of the Russian Ministry of
Industry, Science and Technology No 40.052.1.1.1112 and by the grants:
RFBR 03-02-17373
and the RFBR grant for support of young scientists,
the grant of the Dynasty foundation and MCFFM  (A.A.),
Volkswagen Stiftung (A.Mir. and A.Mor.),
RFBR 04-02-16538, Grant of Support for the Scientific
Schools 96-15-96798 (A.Mir.), RFBR 04-02-16880 (A.Mor.).

\newpage
\setcounter{equation}{0}
\def\theequation{T.\arabic{equation}}

\section*{
\centerline{\Large{
TABLE 1.
}}
\centerline{\Large{
Explicit expressions for the lowest
}}
\centerline{\Large{
$K_W^{(p|m)}$ and  $\check K_W^{(p|m)}$
}}
\label{corext1}}

\be
K_W^{(\cdot|1)}(z;g) = \sum_{p=0}^\infty g^{2p-2}\rho_W^{(p|1)}(z)
\ee

---------------------------------

\be
\check K_W^{(0|1)}(z)=\check\rho_W^{(0|1)}(z) = \frac{W'(z) - \check y(z)}{2}
\label{cheK01}
\ee

\be
K_W^{(0|1)}(z)=\rho_W^{(0|1)}(z) = \frac{W'(z) - y(z)}{2}
\label{K01}
\ee
\be
K_W^{(0|1)}(z)=\left[\check K_W^{(0|1)}\right]^{(0)}(z)
\ee

---------------------------------

\be
\check K_W^{(1|1)}(z) =
-\frac{1}{4\check y^2}\check y^{\prime\prime}
\label{cheK11}
\ee

\be
K_W^{(1|1)}(z) =
-\frac{y^{\prime\prime}}{4y^2} -
\frac{(\check R y)}{2y^2} +
\frac{(\check R F^{(1)})}{y}
\label{K11}
\ee

\be
K_W^{(1|1)}(z)=\left[\check K_W^{(0|1)}\right]^{(1)}(z)+\left[\check K_W^{(1|1)}\right]^{(0)}(z)
\ee

---------------------------------

\be
\check K_W^{(2|1)}(z;g) =
\frac{5}{16}\frac{(\check y^{\prime\prime})^2}{\check y^5} -
\frac{1}{8\check y^2}
\partial^2\left(\frac{\check y^{\prime\prime}}{\check y^2}\right) -
\frac{1}{8}\frac{\check y^{\prime\prime\prime\prime}}{\check y^4}
\label{cheK21}
\ee

\be
K_W^{(2|1)}(z) =
\frac{(\check R F^{(1)})^2+(\check R^2F^{(1)})}{y^3}
-3\frac{(\check R y)(\check R F^{(1)})}{y}+\frac{(\check R F^{(2)})}{y}-\frac{(\check R^2 y)}{y^4}
+\frac{11(\check R y)^2}{4y^5}+
+\frac{2y''(\check R y)}{y^5}-\\-
\frac{(\check R y'')}{y^4}-\frac{y'' (\check R F^{(1)})}{y^4}+\frac{1}{2y^2}\p^2\left(-\frac{(\check R y )}{2y^2}+
\frac{(\check R F^{(1)})}{y}\right)
+\frac{5}{16}\frac{( y'')^2}{ y^5}-\frac{1}{8
y^2}\p^2\left(\frac{y''}{y^2}\right)-\frac{1}{8}\frac{ y ^{(4)}}{ y^4}
\label{K21}
\ee

\be
K_W^{(2|1)}(z)=\left[\check K_W^{(0|1)}\right]^{(2)}(z)+\left[\check K_W^{(1|1)}\right]^{(1)}(z)
+\left[\check K_W^{(2|1)}\right]^{(0)}(z)
\ee
---------------------------------
\be
K_W^{(\cdot|2)}(z_1,z_2;g) =
\sum_{\sigma}^2
\left(\sum_{p_1,p_2=0}^\infty \frac{g^{2p_1+2p_2-4}}{2!(1!)^2}
\rho_W^{(p_1|1)}(z_{\sigma(1)})\rho_W^{(p_2|1)}(z_{\sigma(2)}) +
\sum_{p=0}^\infty \frac{g^{2p-2}}{1!2!}
\rho_W^{(p|2)}(z_{\sigma(1)},z_{\sigma(2)})\right)
\ee
---------------------------
\be
\c K_W^{(0|2)}(z_1,z_2)=:\c \rho_W^{(0|1)}(z_1)\c \rho_W^{(0|1)}(z_2):
=\frac{W'(z_1)W'(z_2)-W'(z_1)\c y(z_2)-W'(z_2)\c y(z_1)+\c y(z_1)\c y(z_2)}{4}
\ee
\be
K_W^{(0|2)}(z_1,z_2)=\rho_W^{(0|1)}(z_1)\rho_W^{(0|1)}(z_2)
=\frac{W'(z_1)W'(z_2)-W'(z_1) y(z_2)-W'(z_2) y(z_1)+ y(z_1) y(z_2)}{4}
\ee
\be
K_W^{(0|2)}(z_1,z_2)=\left[\c K_W^{(0|2)}(z_1,z_2)\right]^{(0)}=
\left[:\check\rho_W^{(0|1)}\check\rho_W^{(0|1)}:\right]^{(0)}(z_1,z_2)
\ee

---------------------------

\be
\c K_W^{(1|2)}(z_1,z_2)=:\c\rho_W^{(0|1)}(z_1)\c\rho_W^{(1|1)}(z_2)+\c\rho_W^{(1|1)}(z_1)\c\rho_W^{(0|1)}(z_2):
+\c \rho_W^{(0|2)}(z_1,z_2)=\nn\\
=-\frac{1}{2\check y(z_1)}\frac{\p}{\p z_2}\frac{\check y(z_1)-\check
y(z_2)}{(z_1-z_2)}
-\frac{W'(z_1)-\c y(z_1)}{2}\frac{1}{4\c y(z_2)^2}\c y(z_2)''-\\
-\frac{W'(z_2)}{2}\frac{1}{4\c y(z_1)^2}\c y(z_1)''
+\frac{1}{4\c y(z_1)^2}\c y(z_1)''\frac{\c y(z_2)}{2}
\ee

\be
K_W^{(1|2)}(z_1,z_2) =
\rho_W^{(1|1)}(z_1)\rho_W^{(0|1)}(z_2) +
\rho_W^{(0|1)}(z_1)\rho_W^{(1|1)}(z_2) +\rho_W^{(0|2)}(z_1,z_2)=\nn\\
=-\frac{1}{2y(z_1)}\left(\frac{\p}{\p z_2}\frac{y(z_1)-y(z_2)}{z_1-z_2}+(\c
R(z_1)y(z_2))\right)+\nn\\
+\left(-\frac{y(z_1)^{\prime\prime}}{4y(z_1)^2} -
\frac{(\check R(z_1) y(z_1))}{2y(z_1)^2} +
\frac{(\check R(z_1) F^{(1)})}{y(z_1)}\right)\frac{W'(z_2) -
y(z_2)}{2}+\nn\\
+\left(-\frac{y(z_2)^{\prime\prime}}{4y(z_2)^2}-
\frac{(\check R(z_2) y(z_2))}{2y(z_2)^2} +
\frac{(\check R(z_2) F^{(1)})}{y(z_2)}\right)\frac{W'(z_1) - y(z_1)}{2}
\ee

\be
K_W^{(1|2)}(z_1,z_2)=\left[\c K_W^{(0|2)}\right]^{(1)}(z_1,z_2)+\left[\c K_W^{(1|2)}\right]^{(0)}(z_1,z_2)=\nn\\
=\left[ \check\rho_W^{(0|2)}\right]^{(0)} (z_1,z_2)
+ \left[:\check\rho_W^{(0|1)}\check\rho_W^{(0|1)}:\right]^{(1)} (z_1,z_2)
+
\left[ :\check\rho_W^{(1|1)}\check\rho_W^{(0|1)} +
 \check\rho_W^{(0|1)}\check\rho_W^{(1|1)}:\right]^{(0)} (z_1,z_2)
\ee

------------------

\be
\c K_W^{(2|2)}(z_1,z_2)=
:\c\rho_W^{(0|1)}(z_1)\c\rho_W^{(2|1)}(z_2)+\c\rho_W^{(0|1)}(z_2)\c\rho_W^{(2|1)}(z_1)
+\c\rho_W^{(1|1)}(z_1)\c\rho_W^{(1|1)}(z_2):
+\c \rho_W^{(1|2)}(z_1,z_2)=\\=
\frac{W'_1-\c y_1}{2}
\left(\frac{5}{16}\frac{(\check y_2'')^2}{\check y_2^5}-\frac{1}{8\check
y_2^2}\p^2\left(\frac{\check y_2''}{\check y_2^2}\right)-\frac{1}{8}\frac{\check y_2 ^{(4)}}{\check
y_2^4}\right)
+\frac{W'_2}{2}
\left(\frac{5}{16}\frac{(\check y_1'')^2}{\check y_1^5}-\frac{1}{8\check
y_1^2}\p^2\left(\frac{\check y_1''}{\check y_1^2}\right)-\frac{1}{8}\frac{\check y_1 ^{(4)}}{\check
y_1^4}\right)-\\
-\left(\frac{5}{16}\frac{(\check y_1'')^2}{\check y_1^5}-\frac{1}{8\check
y_1^2}\p^2\left(\frac{\check y_1''}{\check y_1^2}\right)-\frac{1}{8}\frac{\check y_1 ^{(4)}}{\check
y_1^4}\right)\frac{\c y_2}{2}
+\frac{1}{4\c y_1^2}\c y_1''\frac{1}{4\c y_2^2}\c y_2''+
\frac{1}{\check y_1}\left[
\left(4\frac{1}{4\c y_1^2}\c y_1'' -\frac{1}{2\check y_1}\p_1^2\right)
\frac{1}{2\check y_1}\frac{\p}{\p z_2}\frac{\check y_1-\check
y_2}{(z_1-z_2)}+\right.\nn\\
+\left.\frac{\p }{\p z_2}\frac{1}{z_1-z_2}\left(\frac{1}{4\c y_2^2}\c y_2''-\frac{1}{4\c y_1^2}\c y_1''+
\frac{1}{y_1}\left(-\frac{1}{4\c y_1}\c y''_1+\frac{1}{2\check y_1}\frac{\p}{\p z_2}\frac{\check y_1-\check
y_2}{(z_1-z_2)}\right)
\right]\right)
\ee

\be
K_W^{(2|2)}(z_1,z_2)=\frac{W'_1-y_1}{2}\left(
\frac{(\check R_2 F^{(1)})^2+(\check R_2^2F^{(1)})}{y_2^3}
-3\frac{(\check R_2 y_2)(\check R_2 F^{(1)})}{y_2}+\frac{(\check R_2 F^{(2)})}{y_2}-\frac{(\check R_2^2 y_2)}{y_2^4}
+\right.\\+\frac{11(\check R_2 y_2)^2}{4y_2^5}
+\frac{2y_2''(\check R_2 y_2)}{y_2^5}-
\frac{(\check R_2 y_2'')}{y_2^4}-\frac{y_2'' (\check R_2 F^{(1)})}{y_2^4}+
\frac{1}{2y_2^2}\p_2^2\left(-\frac{(\check R_2 y_2 )}{2y_2^2}+
\frac{(\check R_2 F^{(1)})}{y_2}\right)+\nn\\
+\left.\frac{5}{16}\frac{( y_2'')^2}{ y_2^5}-\frac{1}{8
y_2^2}\p^2\left(\frac{y_2''}{y_2^2}\right)-\frac{1}{8}\frac{ y_2 ^{(4)}}{ y_2^4}
\right)
+\frac{W'_2-y_2}{2}\left(\frac{(\check R_1 F^{(1)})^2+(\check R_1^2F^{(1)})}{y_1^3}
-3\frac{(\check R_1 y_1)(\check R_1 F^{(1)})}{y_1}+
\right.\\\left.+\frac{(\check R_1 F^{(2)})}{y_1}-
\frac{(\check R_1^2 y_1)}{y_1^4}
+\frac{11(\check R_1 y_1)^2}{4y_1^5}
+\frac{2y_1''(\check R_1 y_1)}{y_1^5}-
\frac{(\check R_1 y_1'')}{y_1^4}-\frac{y_1'' (\check R_1 F^{(1)})}{y_1^4}+
\right.\\\left.+
\frac{1}{2y_1^2}\p^2\left(-\frac{(\check R_1 y_1 )}{2y_1^2}+
\frac{(\check R_1 F^{(1)})}{y_1}\right)
+\left.\frac{5}{16}\frac{( y_1'')^2}{ y_1^5}-\frac{1}{8
y_1^2}\p^2\left(\frac{y_1''}{y_1^2}\right)-\frac{1}{8}\frac{ y_1^{(4)}}{ y_1^4}
\right)
+\right.\\\left.+\left(\frac{y_1^{\prime\prime}}{4y_1^2} +
\frac{(\check R_1 y_1)}{2y_1^2} -
\frac{(\check R_1
F^{(1)})}{y_1}\right)\left(\frac{y_2^{\prime\prime}}{4y_2^2}+
\frac{(\check R_2 y_2)}{2y_2^2} -
\frac{(\check R_2 F^{(1)})}{y_2}\right)
+\frac{1}{y_1}\check R_1\left(\frac{(\check R_2 F^{(1)})}{y_2}-\frac{(\check R_2 y_2)}{2
y_2^2}-\frac{y''_2}{4y_2^2}\right)+\right.\\
+\frac{1}{y_1^2}\left(\frac{(\check R_1 F^{(1)})}{y_1}-\frac{(\check R_1
y_1)}{y_1^2}+\frac{\check R_1}{2}\frac{1}{y_1}-\frac{y_1''}{2y_1^4}+\frac{\p_1^2}{4y_1^3}\right)
\left(\p_2 \frac{y_1-y_2}{x_2-x_1}-\check R_1y_2\right)-\nn\\
-\frac{1}{y_1}\p_2\frac{1}{x_1-x_2}\left(\frac{y_2''}{4y_2^2}-\frac{y_1''}{4y_1^2}+
\frac{(\check R_2 y_2)}{2y_2^2}-\frac{(\check R_1 y_1)}{2y_1^2}+\frac{(\check R_1 F^{(1)})}{y_1}
-\frac{(\check R_2 F^{(1)})}{y_2}+\right.\nn\\
+\left.\frac{1}{2y_1}\left((\check R_1y_2)-(\check R_1y_1)+\p_2\frac{y_1-y_2}{x_1-x_2}
-\frac{y_1''}{2}\right)\right)
\ee

\be
K_W^{(2|2)}(z_1,z_2)=\left[\c K^{(0|2)}\right]^{(2)}(z_1,z_2)+\left[\c K_W^{(1|2)}\right]^{(1)}(z_1,z_2)
+\left[\c K_W^{(2|2)}\right]^{(0)}(z_1,z_2)=\nn\\
=\left[ \check\rho_W^{(0|2)}\right]^{(1)} (z_1,z_2)+
\left[ \check\rho_W^{(1|2)}\right]^{(0)} (z_1,z_2)
+ \left[:\check\rho_W^{(0|1)}\check\rho_W^{(0|1)}:\right]^{(2)} (z_1,z_2)
+ \nn \\ +
\left[ :\check\rho_W^{(1|1)}\check\rho_W^{(0|1)} +
 \check\rho_W^{(0|1)}\check\rho_W^{(1|1)}:\right]^{(1)} (z_1,z_2)
+
\left[ :\check\rho_W^{(2|1)}\check\rho_W^{(0|1)} +
\check\rho_W^{(1|1)}\check\rho_W^{(1|1)}+
 \check\rho_W^{(0|1)}\check\rho_W^{(2|1)}:\right]^{(0)} (z_1,z_2)
\ee
---------------------------

\be
K_W^{(\cdot|3)}(z_1,z_2,z_3;g) =
\sum_{\sigma}^6
\left(\sum_{p_1,p_2,p_3=0}^\infty
\frac{g^{2p_1+2p_2+2p_3-6}}{3!(1!)^3}
\rho_W^{(p_1|1)}(z_{\sigma(1)})\rho_W^{(p_2|1)}(z_{\sigma(2)})
\rho_W^{(p_3|1)}(z_{\sigma(3)}) + \right.
\nn \\
\left. +
\sum_{p_1,p_2=0}^\infty \frac{g^{2p_1+2p_2-4}}{(1!)^3 2!}
\rho_W^{(p_1|1)}(z_{\sigma(1)})\rho_W^{(p_2|2)}(z_{\sigma(2)},z_{\sigma(3)}) +
\sum_{p=0}^\infty \frac{g^{2p-2}}{1!3!}
\rho_W^{(p|3)}(z_{\sigma(1)},z_{\sigma(2)},z_{\sigma(3)})\right)
\ee

---------------------------
\be
\c K_W^{(0|3)}(z_1,z_2,z_3)=:\c\rho^{(0|1)}(z_1)\c\rho^{(0|1)}(z_2)
\c\rho^{(0|1)}(z_3):
=\frac{1}{8}\left[W'(z_1)W'(z_2)W'(z_3)-W'(z_1)W'(z_2)\c y(z_3)
-\right.\nn\\
-W'(z_1)W'(z_3)\c y(z_2)-
W'(z_2)W'(z_3)\c y(z_1)+W'(z_1)\c y(z_2) \c y(z_3)
+\nn\\
+\left. W'(z_2)\c y(z_1) \c y(z_3)+W'(z_3)\c y(z_1) \c y(z_2)-\c y(z_1) \c y(z_2)\c y(z_3)
\right]
\ee

\be
\c K_W^{(0|3)}(z_1,z_2,z_3)=\frac{1}{8}\left[W'(z_1)W'(z_2)W'(z_3)-W'(z_1)W'(z_2) y(z_3)
-W'(z_1)W'(z_3) y(z_2)-\right.\\-
W'(z_2)W'(z_3) y(z_1)+W'(z_1) y(z_2)  y(z_3)
+\left. W'(z_2) y(z_1)  y(z_3)+W'(z_3) y(z_1)  y(z_2)- y(z_1)  y(z_2) y(z_3)
\right]
\ee

\be
K_W^{(0|3)}(z_1,z_2,z_3)=\left[\c K_W^{(0|3)}\right]^{(0)}(z_1,z_2,z_3)
=\left[:\c\rho_W^{(0|1)}(z_1)\c\rho_W^{(0|1)}(z_2)\c\rho_W^{(0|1)}(z_3):\right]^{(0)}
\ee
---------------------------
\be
\c K_W^{(1|3)}=\left(:\c\rho_W^{(1|1)}(z_1)\c\rho_W^{(0|1)}(z_2)\c\rho_W^{(0|1)}(z_3):+
:\c\rho_W^{(0|1)}(z_1)\c\rho_W^{(1|1)}(z_2)\c\rho_W^{(0|1)}(z_3):+\right.\nn\\
+:\c\rho_W^{(0|1)}(z_1)\c\rho_W^{(0|1)}(z_2)\c\rho_W^{(1|1)}(z_3):+
:\c\rho_W^{(0|2)}(z_1,z_2)\c\rho_W^{(0|1)}(z_3):+
:\c\rho_W^{(0|2)}(z_1,z_3)\c\rho_W^{(0|1)}(z_2):+\nn\\
+\left.:\c\rho_W^{(0|1)}(z_1)\c\rho_W^{(0|2)}(z_2,z_3):\right)
\ee
---------------------------
\be
\c K_W^{(2|3)}=\left(:\c\rho_W^{(2|1)}(z_1)\c\rho_W^{(0|1)}(z_2)\c\rho_W^{(0|1)}(z_3):+
:\c\rho_W^{(0|1)}(z_1)\c\rho_W^{(2|1)}(z_2)\c\rho_W^{(0|1)}(z_3):+\right.\nn\\
+:\c\rho_W^{(0|1)}(z_1)\c\rho_W^{(0|1)}(z_2)\c\rho_W^{(2|1)}(z_3):+
:\c\rho_W^{(0|1)}(z_1)\c\rho_W^{(1|1)}(z_2)\c\rho_W^{(1|1)}(z_3):+\nn\\
+:\c\rho_W^{(1|1)}(z_1)\c\rho_W^{(0|1)}(z_2)\c\rho_W^{(1|1)}(z_3):+
:\c\rho_W^{(1|1)}(z_1)\c\rho_W^{(1|1)}(z_2)\c\rho_W^{(0|1)}(z_3):+
:\c\rho_W^{(0|2)}(z_1,z_2)\c\rho_W^{(1|1)}(z_3):+\nn\\
+:\c\rho_W^{(0|2)}(z_1,z_3)\c\rho_W^{(1|1)}(z_2):+
:\c\rho_W^{(1|1)}(z_1)\c\rho_W^{(0|2)}(z_2,z_3):+
:\c\rho_W^{(1|2)}(z_1,z_2)\c\rho_W^{(0|1)}(z_3):+\nn\\
+:\c\rho_W^{(1|2)}(z_1,z_3)\c\rho_W^{(0|1)}(z_2):+
:\c\rho_W^{(0|1)}(z_1)\c\rho_W^{(1|2)}(z_2,z_3):+
:\c\rho_W^{(0|2)}(z_1,z_2)\c\rho_W^{(1|1)}(z_3):+\nn\\
+\left.:\c\rho_W^{(0|2)}(z_1,z_3)\c\rho_W^{(1|1)}(z_2):+
:\c\rho_W^{(0|3)}(z_1,z_2,z_3)\right)
\ee
--------------------------
\be
K_W^{(\cdot|4)}(z_1,\ldots,z_4;g) =
\sum_{\sigma}^{4!}
\left(\sum_{p_1,\ldots,p_4=0}^\infty
\frac{g^{2p_1+\ldots+2p_4-8}}{4!(1!)^4}
\rho_W^{(p_1|1)}(z_{\sigma(1)})
\ldots \rho_W^{(p_4|1)}(z_{\sigma(4)}) + \right.
\nn \\
\left. +
\sum_{p_1,p_2,p_3=0}^\infty
\frac{g^{2p_1+2p_2+2p_3-6}}{2!(1!)^21!2!}
\rho_W^{(p_1|1)}(z_{\sigma(1)})\rho_W^{(p_2|1)}(z_{\sigma(2)})
\rho_W^{(p_3|2)}(z_{\sigma(3)},z_{\sigma(4)}) + \right.
\nn \\
\left. +
\sum_{p_1,p_2=0}^\infty \frac{g^{2p_1+2p_2-4}}{(1!)^3 2!}
\rho_W^{(p_1|1)}(z_{\sigma(1)})
\rho_W^{(p_2|3)}(z_{\sigma(2)},z_{\sigma(3)},z_{\sigma(4)}) +
 \right.
\nn \\
\left. +
\sum_{p=0}^\infty \frac{g^{2p-2}}{1!4!}
\rho_W^{(p|4)}(z_{\sigma(1)},\ldots,z_{\sigma(4)}) +
\sum_{p_1,p_2=0}^\infty \frac{g^{2p_1+2p_2-4}}{2!(2!)^2}
\rho_W^{(p_1|2)}(z_{\sigma(1)},z_{\sigma(2)})
\rho_W^{(p_2|2)}(z_{\sigma(3)},z_{\sigma(4)})
\right)
\ee

\newpage

\section*{
\centerline{\Large{
TABLE 2
}}
\centerline{\Large{
Explicit expressions for the lowest
}}
\centerline{\Large{
$\rho_W^{(p|m)}$, $\rho_G^{(p|m)}$ and  $\check\rho_W^{(p|m)}$
}}
\label{corext2}}

---------------------------------

\be
\check\rho_W^{(0|1)}(z;g) = \frac{W'(z) - \check y(z;g)}{2}
\label{cherho01}
\ee
\be
\rho^{(0|1)}_G(z) = \frac{z - y_G(z)}{2}
\label{Garho01}
\ee
\be
\rho_W^{(0|1)}(z) = \frac{W'(z) - y(z)}{2}
\label{rho01}
\ee

---------------------------------

\be
\check\rho_W^{(1|1)}(z;g) =
-\frac{1}{4\check y^2}\check y^{\prime\prime}
\label{cherho11}
\ee
\be
\rho^{(1|1)}_G(z) = \frac{\nu}{y_G^5} =
-\frac{y_G^{\prime\prime}}{4y_G^2}
\label{Garho11}
\ee
\be
\rho_W^{(1|1)}(z) =
-\frac{y^{\prime\prime}}{4y^2} -
\frac{(\check R y)}{2y^2} +
\frac{(\check R F^{(1)})}{y}
\label{rho11}
\ee

---------------------------------

\be
\check\rho_W^{(2|1)}(z;g) =
\frac{5}{16}\frac{(\check y^{\prime\prime})^2}{\check y^5} -
\frac{1}{8\check y^2}
\partial^2\left(\frac{\check y^{\prime\prime}}{\check y^2}\right) -
\frac{1}{8}\frac{\check y^{\prime\prime\prime\prime}}{\check y^4}
\label{cherho21}
\ee
(specification of ordering in this formula requires calculations
with better accuracy),
\be
\rho^{(2|1)}_G(z) =
 \frac{5}{16}\frac{(y_G^{\prime\prime})^2}{y_G^5} -
\frac{1}{8y_G^2}\partial^2\left(\frac{y_G^{\prime\prime}}{y_G^2}\right) -
\frac{1}{8}\frac{y_G^{\prime\prime\prime\prime}}{y_G^4}
\label{Garho21}
\ee

\be
\rho_W^{(2|1)}(z) =
\frac{(\check R F^{(1)})^2+(\check R^2F^{(1)})}{y^3}
-3\frac{(\check R y)(\check R F^{(1)})}{y}+\frac{(\check R F^{(2)})}{y}-\frac{(\check R^2 y)}{y^4}
+\frac{11(\check R y)^2}{4y^5}+\nn\\
+\frac{2y''(\check R y)}{y^5}-
\frac{(\check R y'')}{y^4}-\frac{y'' (\check R F^{(1)})}{y^4}+\frac{1}{2y^2}\p^2\left(-\frac{(\check R y )}{2y^2}+
\frac{(\check R F^{(1)})}{y}\right)+\nn\\
+\frac{5}{16}\frac{( y'')^2}{ y^5}-\frac{1}{8
y^2}\p^2\left(\frac{y''}{y^2}\right)-\frac{1}{8}\frac{ y ^{(4)}}{ y^4}
\label{rho21}
\ee

---------------------------------

\be
\rho_W^{(0|2)}(z_1,z_2) =
\left[ \check\rho_W^{(0|2)}\right]^{(0)} (z_1,z_2)+\nn\\
+\left[:\check\rho_W^{(0|1)}\check\rho_W^{(0|1)}:\right]^{(1)} (z_1,z_2)+
\left[ :\check\rho_W^{(1|1)}\check\rho_W^{(0|1)}+
\left.\check\rho_W^{(0|1)}\check\rho_W^{(1|1)}:\right]^{(0)} (z_1,z_2)\nn\right.\\
-\left[\check\rho_W^{(1|1)}\right]^{(0)}(z_1)\left[\check\rho_W^{(0|1)}\right]^{(0)} (z_2)
-\left[\check\rho_W^{(0|1)}\right]^{(0)}(z_1)\left[\check\rho_W^{(1|1)}\right]^{(0)}
(z_2)-\nn\\
-\left[\check\rho_W^{(0|1)}\right]^{(1)}(z_1)\left[\check\rho_W^{(0|1)}\right]^{(0)} (z_2)
-\left[\check\rho_W^{(0|1)}\right]^{(0)}(z_1)\left[\check\rho_W^{(0|1)}\right]^{(1)} (z_2)
\ee

\be
\check\rho^{(0|2)}(z_1,z_2;g) = -\frac{1}{2\check y(z_1;g)}
\frac{\partial}{\partial z_2}
\frac{\check y(z_1;g) - \check y(z_2;g)}{z_1-z_2}
\label{cherho02}
\ee

\be
\rho^{(0|2)}_G(z_1,z_2) = \frac{1}{2(z_1-z_2)^2}
\left(\frac{z_1z_2-4\nu}{y_G(z_1)y_G(z_2)} -1\right) = \nn \\
 = -\frac{1}{2y_G(z_1)}
\frac{\partial}{\partial z_2}\frac{y_G(z_1) - y_G(z_2)}{z_1-z_2}
 = -\frac{1}{2y_G(z_2)}
\frac{\partial}{\partial z_1}\frac{y_G(z_1) - y_G(z_2)}{z_1-z_2}
\label{Garho02}
\ee

\be
\rho_W^{(0|2)}(z_1,z_2) =
-\frac{1}{2y(z_1)}\left(
\frac{\partial}{\partial z_2}\frac{y(z_1) - y(z_2)}{z_1-z_2}
+ \left(\check R(z_1)y(z_2)\right)\right) = \nn \\ =
-\frac{1}{2y(z_2)}\left(
\frac{\partial}{\partial z_1}\frac{y(z_1) - y(z_2)}{z_1-z_2}
+ \left(\check R(z_1)y(z_2)\right)\right) = \nn \\ =
\frac{W_1'W_2' - y_1y_2 - \left(2(\check R_1 + \check R_2) -
z_{12}(\check R_1' - \check R_2') -
z_{12}^2(\check R_1\check R_2+\check R_2\check R_1)\right)
F^{(0)} }{2z_{12}^2y_1y_2}
\label{rho02}
\ee

---------------------------------

\be
\check\rho_W^{(1|2)}(z_1,z_2;g) =
\frac{1}{\check y_1}\left[
\left(4\frac{1}{4\c y_1^2}\c y_1'' -\frac{1}{2\check y_1}\p_1^2\right)
\frac{1}{2\check y_1}\frac{\p}{\p z_2}\frac{\check y_1-\check
y_2}{(z_1-z_2)}+\right.\nn\\
+\left.\frac{\p }{\p z_2}\frac{1}{z_1-z_2}\left(\frac{1}{4\c y_2^2}\c y_2''-\frac{1}{4\c y_1^2}\c y_1''+
\frac{1}{y_1}\left(-\frac{1}{4\c y_1}\c y''_1+\frac{1}{2\check y_1}\frac{\p}{\p z_2}\frac{\check y_1-\check
y_2}{(z_1-z_2)}\right)
\right]\right)
\label{cherho12}
\ee

\be
\rho^{(1|2)}_G(z_1,z_2) =
\frac{\nu}{y_G^7(z_1)y_G^7(z_2)}
\Big(
z_1z_2(5z_1^4 + 4z_1^3z_2 + 3z_1^2z_2^2 + 4z_1z_2^3 + 5z_2^4) +
 \nn \\  +
4\nu\left[z_1^4 - 13z_1z_2(z_1^2 + z_1z_2 + z_2^2) + z_2^4\right] +
16\nu^2(-z_1^2 + 13z_1z_2 - z_2^2) + 320\nu^3
\Big)=\nn\\
=\frac{1}{ y_{G1}}\left[
\left(4\frac{1}{4 y_{G1}^2} y_{G1}'' -\frac{1}{2 y_{G1}}\p_1^2\right)
\frac{1}{2 y_{G1}}\frac{\p}{\p z_2}\frac{ y_{G1}-
y_{G2}}{(z_1-z_2)}+\right.\nn\\
+\left.\frac{\p }{\p z_2}\frac{1}{z_1-z_2}\left(\frac{1}{4 y_{G2}^2} y_{G2}''-\frac{1}{4 y_{G1}^2} y_{G1}''+
\frac{1}{y_{G1}}\left(-\frac{1}{4 y_{G1}} y''_{G1}+\frac{1}{2 y_{G1}}\frac{\p}{\p z_2}\frac{ y_{G1}-
y_{G2}}{(z_1-z_2)}\right)
\right]\right)
\label{Garho12}
\ee

\be
\rho_W^{(1|2)}(z_1,z_2) =
\frac{1}{y_1}\check R_1\left(\frac{(\check R_2 F^{(1)})}{y_2}-\frac{(\check R_2 y_2)}{2
y_2^2}-\frac{y''_2}{4y_2^2}\right)+\nn\\
+\frac{1}{y_1^2}\left(\frac{(\check R_1 F^{(1)})}{y_1}-\frac{(\check R_1
y_1)}{y_1^2}+\frac{\check R_1}{2}\frac{1}{y_1}-\frac{y_1''}{2y_1^4}+\frac{\p_1^2}{4y_1^3}
\right)\left(\p_2 \frac{y_1-y_2}{x_2-x_1}-\check R_1y_2\right)-\nn\\
-\frac{1}{y_1}\p_2\frac{1}{x_1-x_2}\left(\frac{y_2''}{4y_2^2}-\frac{y_1''}{4y_1^2}+
\frac{(\check R_2 y_2)}{2y_2^2}-\frac{(\check R_1 y_1)}{2y_1^2}+\frac{(\check R_1 F^{(1)})}{y_1}
-\frac{(\check R_2 F^{(1)})}{y_2}+\right.\nn\\
+\left.\frac{1}{2y_1}\left((\check R_1y_2)-(\check R_1y_1)+\p_2\frac{y_1-y_2}{x_1-x_2}
-\frac{y_1''}{2}\right)\right)
\label{rho12}
\ee

---------------------------------

\be
\check\rho^{(0|3)}(z_1,z_2,z_3;g) =
\frac{1}{\c y_1}\left(
2\frac{1}{2\c y_1}\left(\frac{\partial}{\partial z_2}
\frac{\c y_1 -  \c y_2}{z_1-z_2}
\right)\frac{1}{2\c y_1}\left(\frac{\partial}{\partial z_2}
\frac{\c y_1 -  \c y_3}{z_1-z_3}\right)+\right.\nn\\
+\frac{\p}{\p z_2}\frac{1}{z_2-z_1}\left(
\frac{1}{2\c y_1}\left(\frac{\partial}{\partial z_3}
\frac{\c y_1 -  \c y_3}{z_1-z_3}
\right)-
\frac{1}{2\c y_2}\left(\frac{\partial}{\partial z_3}
\frac{\c y_2 -  \c y_3}{z_2-z_3}\right)
\right)+\nn\\
+\left.\frac{\p}{\p z_3}\frac{1}{z_3-z_1}\left(
\frac{1}{2\c y_1}\left(\frac{\partial}{\partial z_2}
\frac{\c y_1 -  \c y_2}{z_1-z_2}
\right)-
\frac{1}{2\c y_3}\left(\frac{\partial}{\partial z_2}
\frac{\c y_2 -  \c y_3}{z_2-z_3}\right)
\right)
\right)
\label{cherho03}
\ee

\be
\rho^{(0|3)}_G(z_1,z_2,z_3) =
\frac{2\nu (z_1z_2 + z_2z_3 + z_3z_1 + 4\nu)
}{y_G^3(z_1)y_G^3(z_2)y_G^3(z_3)}
=\frac{1}{y_{G1}}\left(
2\frac{1}{4y_{G1}^2}\left(\frac{\partial}{\partial z_2}
\frac{y_{G1} -  y_{G2}}{z_1-z_2}
\right)\left(\frac{\partial}{\partial z_2}
\frac{y_{G1} -  y_{G3}}{z_1-z_3}\right)+\right.\nn\\
+\frac{\p}{\p z_2}\frac{1}{z_2-z_1}\left(
\frac{1}{2y_{G1}}\left(\frac{\partial}{\partial z_3}
\frac{y_{G1} -  y_{G3}}{z_1-z_3}
\right)-
\frac{1}{2y_{G2}}\left(\frac{\partial}{\partial z_3}
\frac{y_{G2} -  y_{G3}}{z_2-z_3}\right)
\right)+\nn\\
+\left.\frac{\p}{\p z_3}\frac{1}{z_3-z_1}\left(
\frac{1}{2y_{G1}}\left(\frac{\partial}{\partial z_2}
\frac{y_{G1} -  y_{G2}}{z_1-z_2}
\right)-
\frac{1}{2y_{G3}}\left(\frac{\partial}{\partial z_2}
\frac{y_{G2} -  y_{G3}}{z_2-z_3}\right)
\right)
\right)
\label{Garho03}
\ee

\be
\rho_W^{(0|3)}(z_1,z_2,z_3) = \frac{1}{y_1}\left(
2\frac{1}{4y_1^2}\left(\frac{\partial}{\partial z_2}
\frac{y_1 -  y_2}{z_1-z_2}+(\c R_1y_2)
\right)\left(\frac{\partial}{\partial z_2}
\frac{y_1 -  y_3}{z_1-z_3}+(\c R_1y_3)\right)+\right.-\nn\\
-\left(\c R_1
\frac{1}{2y_2}\left(\frac{\partial}{\partial z_3}
\frac{y_2 -  y_3}{z_2-z_3}+(\c R_2y_3)\right)\right)+\nn\\
+\frac{\p}{\p z_2}\frac{1}{z_2-z_1}\left(
\frac{1}{2y_1}\left(\frac{\partial}{\partial z_3}
\frac{y_1 -  y_3}{z_1-z_3}+(\c R_1y_3)
\right)-
\frac{1}{2y_2}\left(\frac{\partial}{\partial z_3}
\frac{y_2 -  y_3}{z_2-z_3}+(\c R_2y_3)\right)
\right)+\nn\\
+\left.\frac{\p}{\p z_3}\frac{1}{z_3-z_1}\left(
\frac{1}{2y_1}\left(\frac{\partial}{\partial z_2}
\frac{y_1 -  y_2}{z_1-z_2}+(\c R_1y_2)
\right)-
\frac{1}{2y_3}\left(\frac{\partial}{\partial z_2}
\frac{y_2 -  y_3}{z_2-z_3}+(\c R_3y_2)\right)
\right)
\right)
\label{rho03}
\ee

\end{document}